\documentclass[useAMS,usenatbib]{mn2e}
\usepackage{graphicx}
\title[New Brown Dwarf Disks In Upper Scorpius]{New brown dwarf disks in Upper Scorpius observed with WISE}
\author[P. Dawson,  A. Scholz, T.P.Ray, K.A. Marsh, K. Wood, A. Natta, 
D. Padgett, M.E. Ressler.]{P. Dawson,$^{1}$\thanks{E-mail:
pdawson@cp.dias.ie (PD); aleks@cp.dias.ie (AS); tr@cp.dias.ie (TR); ken.marsh@astro.cf.ac.uk (KM); kw25@st-andrews.ac.uk (KW); 
natta@arcetri.astro.it (AN); dlp@ipac.caltech.edu (DP); michael.e.ressler@jpl.nasa.gov (MR)} A. Scholz,$^{1}$ T.P. Ray,$^{1}$ K.A. Marsh,$^{2}$ 
K. Wood,$^{3}$ A. Natta,$^{1,4}$
\newauthor
D. Padgett,$^{5}$ M.E. Ressler.$^{6}$\footnotemark[1]\\
$^{1}$School of Cosmic Physics, Dublin Institute for Advanced Studies, 31 Fitzwilliam Place, 
Dublin 2, Ireland \\$^{2}$School of Physics and Astronomy, Cardiff University, Cardiff CF24 3AA, UK 
\\$^{3}$School of Physics and Astronomy, University of St. Andrews, North Haugh, St. Andrews KY16 9SS, UK 
\\$^{4}$INAF - Osservatorio Astrofisico di Arcetri, Largo E. Fermi 5, 50125 Firenze, Italy
\\$^{5}$Goddard Space Flight Center, Greenbelt, MD 20771, USA
\\$^{6}$Jet Propulsion Laboratory, California Institute of Technology, 4800 Oak Grove Drive, Pasadena, CA 91109, USA}
\begin{document}

\date{Accepted 2012 xxxxxxx xx. Received 2012 xxxxxxx xx; in original form 2012 October xx}

\pagerange{\pageref{firstpage}--\pageref{lastpage}} \pubyear{2009}

\maketitle

\label{firstpage}

\begin{abstract}

We present a census of the disk population for UKIDSS selected brown dwarfs in the 5-10 Myr old Upper Scorpius OB association.   For 116 
objects originally identified in UKIDSS, the majority of them not studied in previous publications, we obtain photometry from 
the WISE database.   The resulting colour-magnitude and colour-colour plots clearly show two separate populations of objects, interpreted 
as brown dwarfs with disks (class II) and without disks (class III).   We identify 27 class II brown dwarfs, 14 of them not previously known.   
This disk fraction (27 out of 116 or 23\%) among brown dwarfs was found to be similar to results for K/M stars in Upper Scorpius, 
suggesting that the lifetimes of disks 
are independent of the mass of the central object for low-mass stars and brown dwarfs.   5 out of 27 disks (19\%) lack excess at 3.4 and 
4.6\,$\mu$m and are potential transition disks (i.e. are in transition from class II to class III).   The transition disk fraction is comparable 
to low-mass stars.   We estimate that the timescale for a typical transition from class II to class III is less than 0.4\,Myr for brown 
dwarfs.   These results suggest that the evolution of brown dwarf disks mirrors the behaviour of disks around low-mass stars, with disk 
lifetimes on the order of 5-10\,Myr and a disk clearing timescale significantly shorter than 1\,Myr.

\end{abstract}


\section{Introduction}
Brown dwarfs -- substellar objects with masses below the Hydrogen burning limit of 0.08$\,M_{\sun}$ --
are ideal to test the mass dependence of critical parameters in stellar evolution. One example for
such a parameter is the lifetime of circumstellar disks, which is an important constraint for
the core-accretion scenarios for planet formation. The disk lifetime is affected by a number of
physical processes, e.g. disk ionisation by the central object and cosmic rays, accretion, grain growth (e.g. 
\citet{dul07}). Our understanding of the relative importance of these processes 
and how they change with object mass is still incomplete, i.e. observational guidance is important to 
advance the theory.

The clear majority of low-mass stars lose their disk within less than 5\,Myr \citep{hai01, jay06}. 
Maybe the best test for the longevity of disks 
is the Upper Scorpius OB assocation (UpSco in the following), the oldest nearby star forming region 
with a substantial number of brown dwarfs. 
UpSco is often assumed to have an age of 5\,Myr \citep{pre02} but recently \citet{pec12} have derived an older age of 10\,Myr for this region.   
In UpSco, \citet{car06} derived a disk frequency of $<8$\% 
for F and G stars and $19\pm4$\% for K0-M5 stars (with 1\,$\sigma$ binomial confidence intervals). The brown dwarfs 
in this region exhibit a disk fraction of $37\pm 9$\%, based on an examination of 35 objects \citep{sch07}. Thus, the Spitzer data tentatively 
shows that the disk fractions {\it in UpSco increase monotonically with decreasing object mass for early F- to late 
M-type objects}.   This would imply a mass dependence in the disk evolution, resulting in long-lived disks in the 
substellar regime.

So far, the brown dwarf disk frequency in UpSco is affected by low number statistics.   
Here we set out to test previous findings 
based on a much enlarged number of brown dwarfs in UpSco, identified from UKIDSS. To identify the objects 
in our sample that have a disk (class II objects) and those that do not (class III objects), 
we use data from the Wide-Field Infrared Survey Explorer (WISE)\citep{wri10}. As will be shown, with improved 
statistics we find a disk fraction for brown dwarfs that is consistent with the value published for low-mass
stars in this region. 

\begin{figure*}

\begin{tabular}{|c|c}

\includegraphics[scale=0.64]{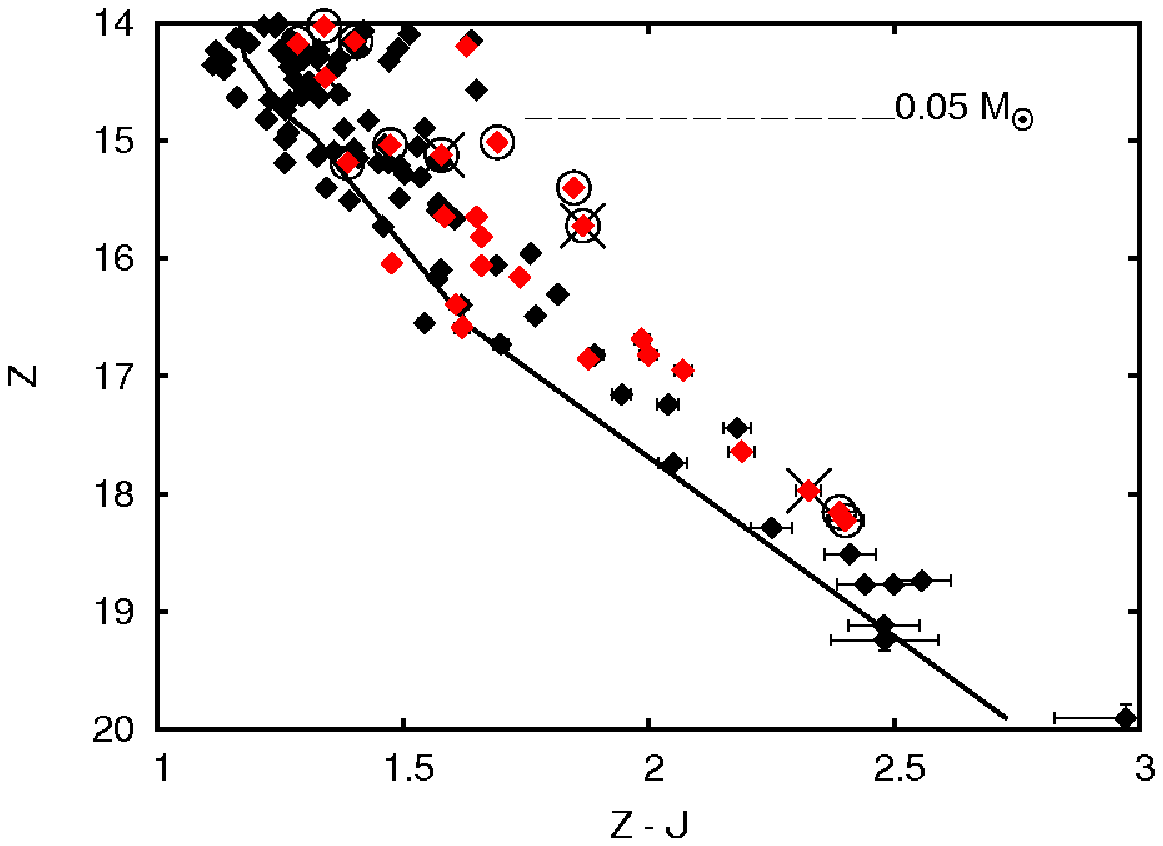}
\includegraphics[scale=0.64]{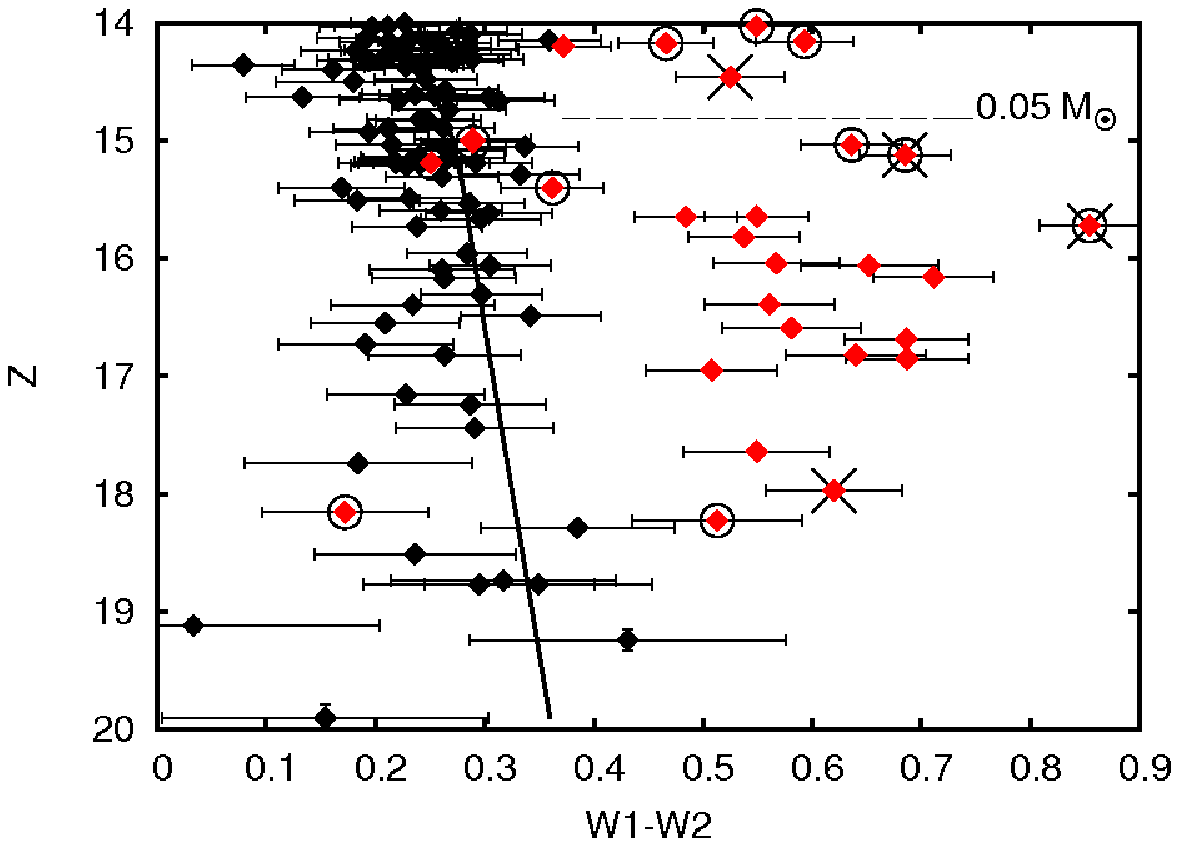}  \\

\end{tabular}
  \caption{(Z-J,Z) and (W1-W2,Z) colour-magnitude diagrams for 116 out of 119 brown dwarfs identified in UpSco 
for which WISE data were available.   The class II objects are marked in red in the online version of this paper.   
The 5 Myr DUSTY model \citep{cbah00} isochrones 
are also shown with mass decreasing 
from 0.09$\,M_{\sun}$ (top) to 0.01$\,M_{\sun}$ (bottom).   
The 0.05$\,M_{\sun}$ positions on the isochrones are indicated.   In the (Z-J,Z) diagram all the objects are grouped close to the 
isochrone and none show any significant colour excess.   However, two distinct populations of objects are clearly visible in the (W1-W2,Z) 
diagram, one close to the isochrone and one showing an excess in W1-W2.   The 11 objects ringed have bright unambiguous signals in W4 (22\,$\mu$m), 
diagnostic of the presence of a disk (see text; section 4.1.4).   The 4 objects marked with a cross have significant variations in their UKIRT and 
2MASS J or H magnitudes, a characteristic associated with accretion events or variable extinction (also see text; section 4.2).}
\end{figure*}

\begin{figure*}

\begin{tabular}{|c|c}

\includegraphics[scale=0.64]{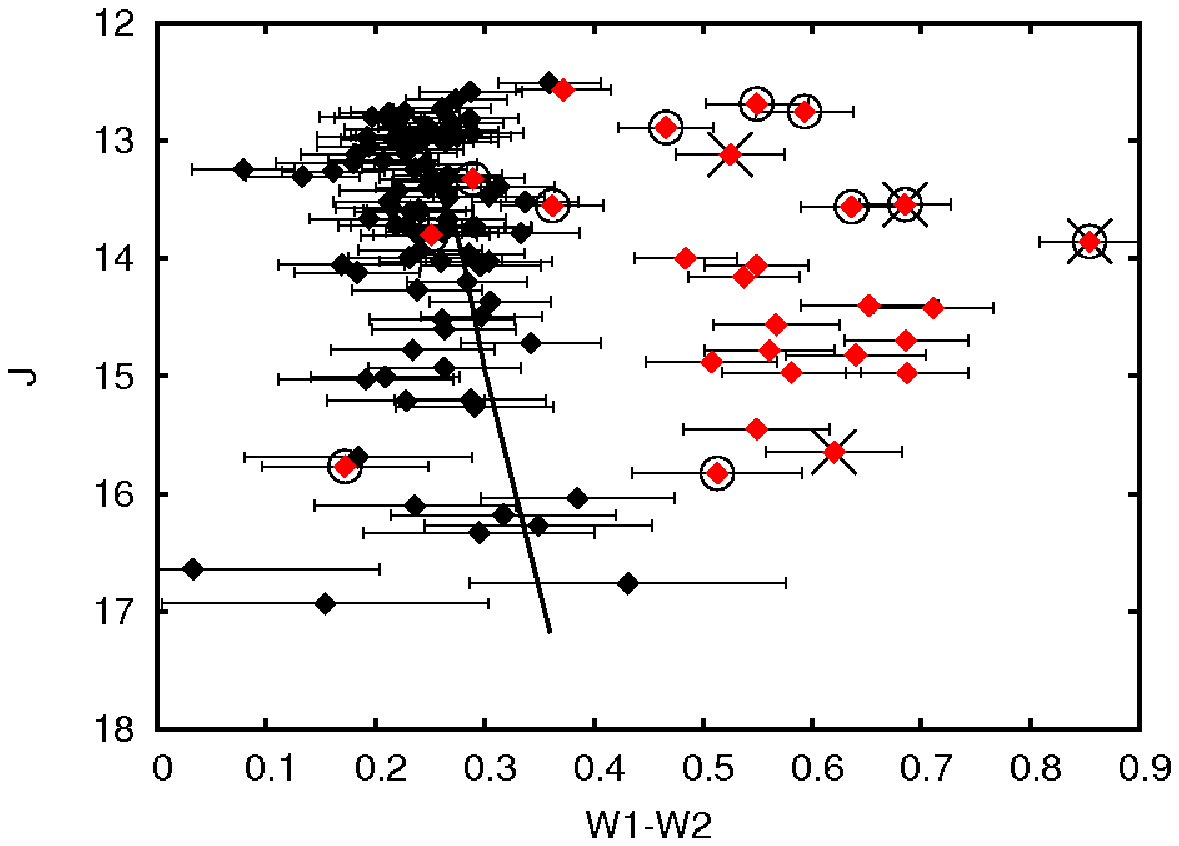}
\includegraphics[scale=0.64]{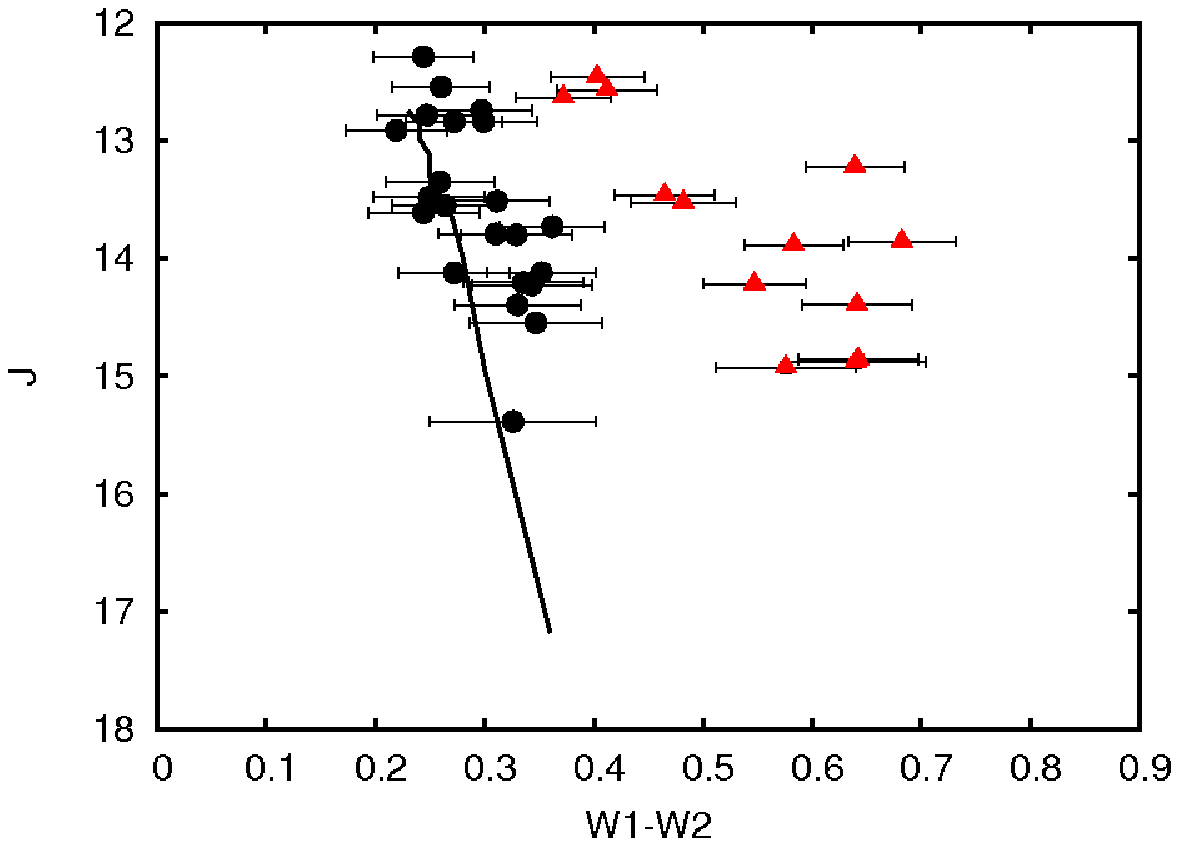}  \\

\end{tabular}
  \caption{(W1-W2,J) colour-magnitude diagrams.   The 116 brown dwarfs with available WISE data 
analysed in this work are shown in the left panel with symbols 
and isochrone as in figure 1.   The right panel shows the same diagram for 35 objects in UpSco identified in \citet{sch07}.   Objects 
confirmed as having disks in \citet{sch07} are shown as closed triangles while those with no disk detected are shown as closed circles.   
As can be seen, the objects in the population with obvious colour excess all have disks.   
A few objects with disks exhibit a smaller, but still significant colour excess (i.e. they are more than 2\,$\sigma$ away from the isochrone).   
Objects with no disk lie close to the isochrone.}
\end{figure*}

\section[]{Targets}
\citet{dsr11} identified 19 new brown dwarf candidates in the south of UpSco via a photometric and proper motion analysis of UKIDSS data.   
The level of contamination from background stars in the sample was shown to be negligible.   
Using the same method in the north of UpSco a further 49 objects previously identified by \citet{lodieu06, lodieu07} as brown dwarf candidates 
were recovered.   
Spectra have been taken of 26 of these 49 objects \citep{mar04, sle06, lodieu06, lodieu08, lodieu11} and all 26 have been confirmed as 
brown dwarfs.   This provides a sample of 68 very low mass stellar and substellar objects in UpSco.   
\citet{dsr11} determined that the 68 objects ranged in mass from 0.01$\,M_{\sun}$ to 0.09$\,M_{\sun}$ by comparing their observed Z magnitudes 
with theoretical Z magnitudes from the DUSTY models for 5 Myr old objects.   

\citet{dsr11} used the 8th Data Release from UKIDSS to identify the new brown dwarfs.   Since then, UKIDSS has issued a 9th Data Release 
which covers a substantially larger area in UpSco.   Using the same method as in \citet{dsr11} we have identified a further 51 objects 
in the same mass range to add to 
the sample.   The details of this extended survey are given in Appendix A.   The addition of these new objects increases the size of the 
homogeneous sample of objects uniformly selected via their colour and proper motion characteristics to 119.   Although some objects in our 
sample may be slightly above the substellar threshold, we will refer to our targets as ``brown dwarfs'' throughout this paper, for simplicity.   

All 119 objects lie close to the 5 Myr DUSTY model isochrones of \citet{cbah00} in a (Z-J,Z) UKIDSS passbands colour-magnitude 
diagram, as shown in figure 1, i.e. none of them exhibit any substantial excess at these near infrared wavelengths.   
The objects are also generally free from reddening caused by extinction (\citet{dsr11}, also see Appendix A this paper).   
Consequently it can be inferred that the (Z-J,Z) value for any of the objects is photospheric in origin with negligible contribution from 
any circum-substellar disk.   Thus, the method of selecting these 119 objects is unbiased with respect to the presence of circum-substellar 
disks.   

As pointed out above, the 26 objects in our sample with published spectra have been confirmed to be very low mass members of UpSco. 
In addition, we have recently obtained spectra for 25 further objects from this sample; and all of them are confirmed as very low mass members 
of UpSco as well (Dawson et al., in preparation). Although the spectroscopic follow-up is not yet complete for our sample, the 100\% success rate 
for almost half the sample indicates that our selection method based on photometry and proper motion generates a clean, unbiased sample 
with negligible contamination by background objects ($\le 2$\%).

\begin{table*}
 \begin{minipage}{170mm}
  \caption{Objects with photometric variability $>$\,0.2mag in J,H or K.}
  \begin{tabular}{|c|c|c|c}
  \hline
    Name & $\Delta$J & $\Delta$H & $\Delta$K\\

 \hline
2MASSJ15514709-2113234 & 0.84 & 0.59 & 0.25\\
2MASSJ15521088-2125372 & 0.24 & 0.11 & 0.03\\
2MASSJ15472282-2139141 & 0.38 & 0.25 & 0.21\\
2MASSJ16030235-2626163 & 0.17 & 0.30 & 0.11\\

\hline

\end{tabular}
\end{minipage}
\end{table*}

\section{WISE Data}

For 116 of the 119 targets discussed above we obtained WISE data.   The remaining 3 were not listed in the WISE database.   
For the 116 the 2MASS identifier listed in the WISE database agrees with the one from UKIDSS.   In addition, all objects were visually 
examined in the WISE images and in the UKIDSS images.   It was found that for the 116 objects analysed here the identification of the 
UKIDSS source with the WISE source at the same position is unambiguous (for the three excluded targets this is not the case).   Furthermore,
since our target fields are sparsely populated and well above the Galactic plane (latitude 10-30\,deg), the likelihood 
of accidental contamination by other objects is very low.    

WISE surveyed the whole sky in four mid-infrared wavebands simultaneously, 
using passbands with effective wavelengths of 3.4\,$\mu$m (W1), 4.6\,$\mu$m (W2), 12\,$\mu$m (W3) and 22\,$\mu$m (W4).   We used the results of 
profile-fitting photometry from the All Sky Data Release, further details of which can be found in \citet{cut12}\footnotemark[1]\footnotetext[1]
{For more details of the WISE All Sky Data Release also see http://wise2.ipac.caltech.edu/docs/release/allsky/expsup/}.
Among the 116 objects are 2 that were previously examined by \citet{sch07} and confirmed to have disks.

\section{Colour Analysis}

The five UKIRT passbands; Z, Y, J, H and K have respective wavelengths of 0.88\,$\mu$m, 1.03\,$\mu$m, 1.25\,$\mu$m, 1.63\,$\mu$m, 
and 2.20\,$\mu$m.   Similar J, H and K passbands are also used in 2MASS.   As noted above, the four WISE passbands; W1, W2, W3, W4, have longer 
wavelengths of 3.4\,$\mu$m, 4.6\,$\mu$m, 12\,$\mu$m and 22\,$\mu$m respectively.    Different colour-colour and colour-magnitude diagrams using 
combinations of all nine passbands were examined.  The (Z-J,Z) and (W1-W2,Z) colour-magnitude diagrams are shown in figure 1.   
Theoretical isochrones for 5 Myr old sub-stellar objects are also shown over-plotted on the diagrams.   
These isochrones are based on the DUSTY models derived by \citet{cbah00} and obtained from both I. Baraffe and 
F. Allard (private communications).   The isochrones were computed using both the UKIDSS and WISE 
filter profiles.   The (Z-J,Z) isochrone was used to assign masses to objects by \citet{dsr11}.   The uppermost point on the 5 Myr isochrone 
corresponds to a mass of 0.09$\,M_{\sun}$ while the lowest point corresponds to a mass of 0.01$\,M_{\sun}$.   

\subsection{WISE data for UpSco brown dwarfs}
\subsubsection{W1-W2 (3.4\,$\mu$m-4.6\,$\mu$m)}
In the (Z-J,Z) diagram all the objects are grouped close to the isochrone and none show any significant colour excess, as discussed above.   
However, two distinct populations of objects are clearly visible in the (W1-W2,Z) diagram, one close to the isochrone and one showing an excess in 
W1-W2 (3.4\,$\mu$m-4.6\,$\mu$m).    
The population with excess is best understood as objects harbouring dusty disks and therefore emitting thermal infrared radiation (class II).   
These two populations do not show up in the diagrams that utilise only UKIRT passbands.   
Other diagrams that combine both UKIRT and WISE passbands do show the two populations.   Colours that use a UKIRT and 
WISE W1 (3.4\,$\mu$m) passband do show the two populations but not as clearly as colours incorporating W2 (4.6\,$\mu$m).   The K-W2 colour in particular 
discriminates between the two populations almost as well as the W1-W2 colour.   
The W1-W2 colour was chosen as the primary diagnostic for distinguishing between class II and class III objects in this work (see figures 1 and 2).   

\subsubsection{W1-W2: Comparison with Scholz (2007)}

To test that the method outlined above was successfully discriminating between class II and class III objects it was also applied to the 35 objects 
in UpSco examined by \citet{sch07}.    That study used a Spitzer survey combining spectroscopy from 8 to 12\,$\mu$m and photometry 
at 24\,$\mu$m.   As 33 of the 35 objects lie outside the area covered by UKIDSS there is no Z or Y passband data available for them.   However 
they are recorded in the J, H and K passbands of 2MASS.   The 35 objects were plotted in a (W1-W2,J) colour-magnitude diagram (shown in the right 
panel of figure 2) using data from 2MASS.   The 116 objects from this work were also plotted in a (W1-W2,J) colour-magnitude diagram (shown in 
the left panel of figure 2) for comparison, using data from the UKIDSS J passband.   Two of the objects from \citet{sch07} lie inside 
the area covered by UKIDSS and were recovered by \citet{dsr11}.   The differences in their UKIDSS and 2MASS J magnitudes are negligible (0.06mag in 
both cases).   

As can be seen from the right panel of figure 2, all the objects with circum-substellar disks have sufficient excess in W1-W2 (3.4\,$\mu$m-4.6\,$\mu$m) 
to stand clear of the objects with no disks clustered along the isochrone.   A few of the objects with disks exhibit a lesser, but still significant 
colour excess in W1-W2 (i.e. they are more than 2\,$\sigma$ away from the isochrone).   The method successfully discriminated between the 13 class II 
and 22 class III objects in this sample.

\subsubsection{W3 (12\,$\mu$m)}
Colours that utilise the longer wavelength W3 (12\,$\mu$m) passband are of limited use as only 39 of the 116 objects have a S/N $>$\,5.0 in W3.   
By contrast, all 116 objects have a S/N $>$\,8.0 in W1 and W2.   
The 39 objects with a S/N $>$\,5.0 in W3 were further examined in the (W1-W2,W3) colour-magnitude diagram shown in figure 3.   
They are all among the higher mass objects in the sample, as evidenced by the lack of objects around the lower 
part of the isochrone.   
The population with the W1-W2 
(3.4\,$\mu$m-4.6\,$\mu$m) colour excess 
is again distinct from the population close to the isochrone as in the previous diagrams.   
The ninth brightest 
object in W3 now stands clear from the population near the isochrone.   While this object is not obviously part of the population with excess in 
W1-W2 seen in figure 1 and figure 2, it is more than 2\,$\sigma$ away from the isochrone.   On the combined basis of its 
excess in W1-W2 and its brightness in W3 it appears to have a disk and so is included in the group of class II objects.   
This object is one of the 2 objects common to both this study and that of \citet{sch07} who note that it has a binary 
companion at a separation of 12 au.   


\begin{figure}
  \includegraphics[width=0.45\textwidth]{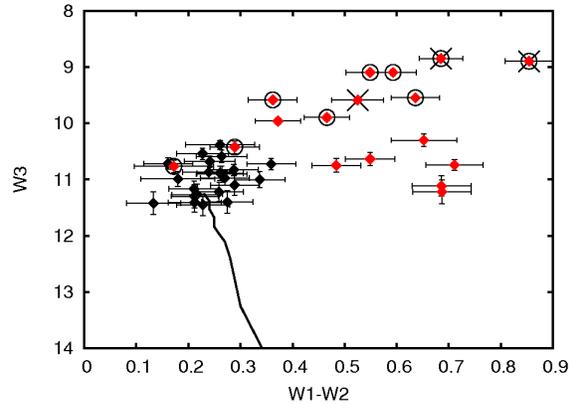}
  \caption{(W1-W2,W3) colour-magnitude diagram for 39 of the 116 objects with a S/N $>$\,5.0 in W3 (12\,$\mu$m).   Symbols and isochrone are 
as in figure 1.   
The 39 objects are all among the higher mass objects in the sample, as evidenced by the lack of objects around the lower part of the isochrone.}
\end{figure}

\subsubsection{W4 (22\,$\mu$m)}

The objects also had their detection in the W4 (22\,$\mu$m) WISE passband examined.  Photospheric emission from brown dwarfs is negligible by comparison 
with emission from a disk at wavelengths longer than 20\,$\mu$m \citep{sch07}.   Therefore any object with a bright unambiguous signal in W4 shows 
clear evidence of the presence of a dusty disk, even if it does not have an excess in W1-W2 (3.4\,$\mu$m-4.6\,$\mu$m).   
105 of the objects were not distinguishable from the background in W4, having S/N varying 
from 4.7 to zero.   The remaining 11 were detected with S/N $>$\,5.0 and have the brightest signals in W4.   These 11 are 
marked with rings in figures 1, 2, 3 and 4.   Of the 11, 7 are in the population of 22 with distinct W1-W2 (3.4\,$\mu$m-4.6\,$\mu$m) colour excess.

\subsubsection{(W1-W2, J-K)}

The (W1-W2,J-K) colour-colour diagram in figure 4 was examined to see if class II and class III objects could be clearly separated.   Apart 
from the 2 objects which are bright in W4 and have large differences in their UKIDSS and 2MASS J magnitudes, none of the objects show 
any significant excess in J-K.   This serves to confirm previous findings \citep{nat01, nat02} that the efficacy of 
using J-K excess as a diagnostic for the presence of disks around very low mass stars and brown dwarfs is very limited.

\begin{figure}

\includegraphics[scale=0.64]{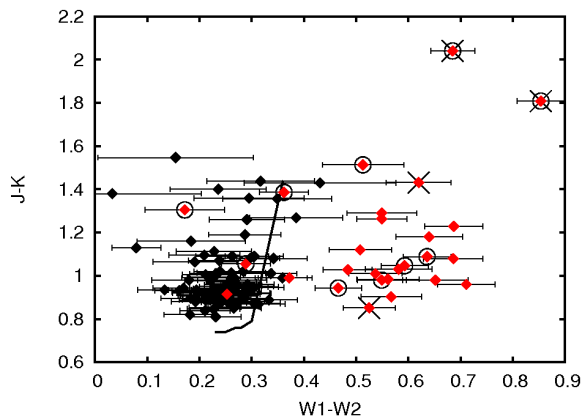}

\caption{(W1-W2,J-K) colour-colour diagram with brown dwarfs and isochrones as in figure 1 and figure 2.   The two objects which 
are bright in W4 and which show signs of accretion have an excess in both W1-W2 and J-K.   Apart from those two the rest of the population 
with excess in W1-W2 shows no obvious excess in J-K.}
\end{figure}

\subsection{Variable Objects}

A comparison was also made of the UKIDSS and 2MASS photometry for each object.   UKIDSS and 2MASS data were gathered at different epochs several 
years apart - 2MASS between 1997 and 2001 \citep{skrutskie06} and UKIDSS from 2005 onwards \citep{lawrence07}.   
Of the 116 objects, 112 showed a variation of less than 0.2 mag in J, H and K.   The remaining 4, which are listed in Table 1 (and marked 
with crosses in figures 1, 2, 3 and 4), showed variations of greater than 0.2 mag in J, H or K.   Such variations in J have been interpreted by 
\citet{sch09a} as 
signatures of accretion.   \citet{sch09a} further note that cool spots (comparable to sunspots) are expected to produce variations 
of $<$\,0.15mag in J and $<$\,0.1mag in K while large-scale photometric variability with amplitudes declining towards longer wavelengths - as 
displayed by the 3 most variable objects in Table 1 - is generally caused by hotspots or variable extinction due to a rotating disk.   Hotspots in 
young stars and brown dwarfs are thought to be a direct consequence of accretion and so they are evidence of the existence of a disk.   
Likewise, variable extinction is also evidence of the existence of a disk.   All 4 objects are 
in the population with W1-W2 colour excess.

\section{Discussion}
In all, 22 objects were identified as class II objects on the basis of their W1-W2 colour alone.   
1 object with a small W1-W2 excess was also placed in the population of class II objects because of its bright W3 signal.   
A further 4 other objects with 
no W1-W2 excess were also categorised as class II because of their bright signals in the 22\,$\mu$m W4 passband.   All 27 objects 
are listed in Table 2.     
The remaining 89 objects, which are listed in Table 3, were deemed to be class III i.e. 
they have no disks or disks with a large inner opacity hole of at least 5 - 20AU (see \citet{sch07}).   
We show the spectral energy distributions for 6 characteristic examples of the class II objects in Appendix B.

\subsection{Disk Fraction}

The overall disk fraction is 27 out of 116 or 
23\%$\pm5$\% (the uncertainty corresponds to a 1\,$\sigma$ confidence interval based on binomial statistics).   
As noted in section 2 above, contamination in our sample appears to be negligible.   Conservatively assuming that 10\% of the objects are 
contaminants which do not exhibit mid-infrared excess would only increase the disk fraction to 27/104, i.e. 26\%.   
This contrasts with the 
previous results for UpSco of $37\pm9$\% \citep{sch07} obtained with a smaller sample of 35 objects.   
The sample of 35 objects in \citet{sch07} was selected from the surveys of \citet{ard00} (12 objects) and \citet{mar04} (23 objects).   
The higher disk fraction reported in \citet{sch07} may be the result of using a smaller sample of objects or of a possible bias.   


The results of \citet{sch07} were also in contrast to the disk fraction of 19\% for K0 to M5 stars in the same region obtained by \citet{car06} 
using a sample of 127 K0 to M5 stars.   
Our new result, of 23\%$\pm5$\%, from a similarly sized sample of 116 objects, is statistically indistinguishable from the result of \citet{car06}, 
suggesting that disk lifetimes in UpSco for objects later than K0 show no dependency on the mass of the central object.   
We note that from our sample of 27 disks, 22 (from 72, 31\%) are found to be in the mass range 0.01-0.05$\,M_{\sun}$, while only 5 (from 44, 11\%) 
are in the mass range 0.05-0.09$\,M_{\sun}$.   This may indicate a trend towards higher disk fractions for very low mass brown dwarfs, but it 
is not sufficiently robust to warrant further discussion.   

\subsubsection{Disk Fractions In Other Clusters}

This result for UpSco can be compared with those for Cha I \citep{dam07, luh05}, IC348 \citep{luh05, lad06} and 
$\sigma$\,Ori \citep{her07}, 3 other associations for which similar information is available.   
\citet{dam07} reports a disk fraction of 52\%$\pm6$\% from a sample of 81 K3 to M8 objects in Cha I, while \citet{luh05} finds a disk fraction 
of 50\%$\pm17$\% from a much smaller sample of 18 objects later than M6.   \citet{luh05} also finds a disk fraction of 42\%$\pm13$\% for 24 
objects later than M6 in IC348, where \citet{lad06} reports a disk fraction of 47\%$\pm12$\% in the range of K6 to M2 stars.   In an analysis 
of disks in $\sigma$\,Ori, \citet{her07} finds a disk fraction of 36\%$\pm4$\% for stars in the mass range of 0.1 to 1.0$\,M_{\sun}$ 
and 33\%$\pm10$\% for brown dwarfs (defined as objects $<$\,0.1$\,M_{\sun}$).   The spectral type of the most massive stars in this range defined 
by \citet{her07} are a little earlier than K (up to G8).   However, the vast bulk of their sample is in the mass range of K and M stars, so 
their result is noted here as being valid for K and M stars in $\sigma$\,Ori.   
Figure 5 shows that in the wake of this revision of the results for UpSco, all 4 associations now show similar disk fractions for K/M stars 
and brown dwarfs.   Thus, average disk lifetime does not appear to be dependent on the mass of the central object in any of these regions.   

\subsubsection{The Ages Of The Associations}

All the associations listed above are young, i.e. $<$\,10 Myr old.   
\citet{pre02} determined an age of about 5 Myr for UpSco, while a revised age of 10 Myr has been 
proposed by \citet{pec12}.   Cha I, IC348 and $\sigma$\,Ori have each had various ages of between 2 and 5 Myr reported for each of them 
\citep{luh04a, luh07a, luh03, may07, zap02, olv02, she04}.   So while UpSco appears to be the oldest of the 4 associations, Cha I, IC348 
and $\sigma$\,Ori cannot yet be readily 
distinguished in terms of their ages.   Ergo, apart from stating that the oldest association has the smallest disk fraction, no robust 
correlation can be safely determined in respect of the disk fractions and ages of the 4 associations.

\begin{figure}
  \includegraphics[width=0.45\textwidth]{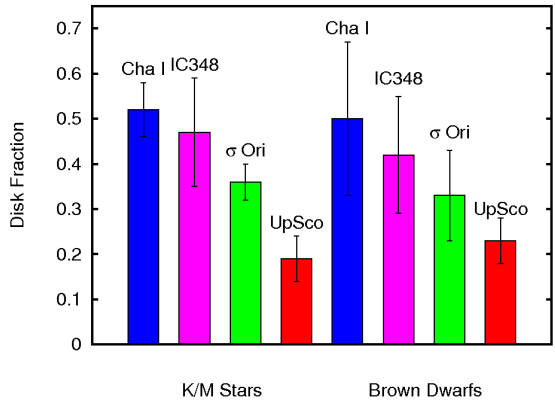}
  \caption{Disk fractions for K/M stars compared to disk fractions for brown dwarfs in Cha 1, IC348, $\sigma$\,Ori and UpSco.   In each case 
the disk fraction for the K/M stars is statistically indistinguishable from the disk fraction for the brown dwarfs (see text; section 5.1).}
\end{figure}





\subsection{Transition disks}

Objects with very little or no excess at near- to mid-infrared wavelengths but exhibiting excess at longer wavelengths are best 
understood by assuming an opacity hole in the inner disk (see section 5.3).   They may be in the process of clearing dust from their inner disks 
\citep{cal02, muz06}.   Such objects are often termed 'transition disks', but the criteria used to define transition 
disks differ in the literature (see \citet{lam12, mer10} for a discussion of the various criteria).   
Here we use the term to identify objects with little or no excess in W1-W2, i.e. shortwards of $10\,\mu$m, but bright W3 and/or W4 
signals (see Sect. 4.1.3 and 4.1.4).   The five objects that satisfy this criterion are included in the class II group and may be in the 
process of a transition from class II to class III.   

We do not find any object that might be called a 'pre-transitional disk' \citep{esp08, esp12}, showing evidence for an opacity 
gap (as opposed to an inner hole) in the disk, i.e. with exess at $<\,10\,\mu$m, no excess at 12$\,\mu$m, and excess again at 22$\,\mu$m.   

Based on our adopted definition, the fraction of transition disks around class II brown dwarfs in our sample is 5/27 or 19\%.   Due to 
the small sample size, the uncertainty in this number is in the range of $\pm10$\%.   Taking this into account, the value is consistent with 
most previous estimates for the transition disk fraction for low-mass stars which are, for criteria similar to the one adopted here, in 
the range of 0-20\% (e.g., \citet{ecr09, muz10}).   Thus, based on our estimate for the brown dwarf regime,
there is no evidence for a mass-dependence in the transition disk frequency.

The small number of transition disks in our sample indicates that the transition phase lasts only a short time compared with the total 
lifetime of the disks.   Assuming the upper limit for the age spread of 2\,Myr \citep{paz99}, we obtain an upper limit of 
0.4\,Myr for the transition timescale, i.e. about one order of magnitude shorter than the disk lifetime.   Thus a two timescale model for the 
evolution of the disks, as often adopted for low-mass stars, is required for brown dwarfs as well.

\subsection{Radiative Transfer Models}
\citet{sch07} produced model spectral energy distributions based on Monte Carlo radiative transfer simulations for the 13 class II objects 
that they found in UpSco.   Figure 6 shows the spectral energy 
distributions of the 2 of those objects recovered in this work, which have been recreated using the original model parameters.   Data shown 
for the J, H, K, 9, 10, 11 and 24\,$\mu$m wavelengths is taken from that of \citet{sch07}.   The new data 
for the W1, W2 and W3 passbands is also shown overplotted on the two spectral energy distribution diagrams.   It is clear from figure 6 that 
the data for W1 (3.4\,$\mu$m), W2 (4.6\,$\mu$m) and W3 (12\,$\mu$m) agree very well with both original models, lying on or very close to 
the modelled flux for a combined photosphere and disk (solid lines in figure 6).   

Models of the inner part of a disk rely on observational data in the mid infrared to refine their accuracy.   
\citet{sch07} noted that the gap in their data coverage in the 3-8\,$\mu$m region 
restricted their ability to constrain the size of any inner disk holes in sources that exhibited excesses at 9\,$\mu$m and beyond.   
The addition of the W1 and W2 datapoints for the two objects in figure 6 now allows the accuracy of the models in this region to be probed.   

The data for the object in the left panel was originally fitted with a model which included an excess in the mid infrared which 
necessitated the presence of an optically thick inner disk.   The newly overplotted W1 (3.4\,$\mu$m) and W2 (4.6\,$\mu$m) values observed 
by WISE conform with that part of the model, further indication that a significant W1-W2 excess is evidence for the presence of a disk.   
The W1-W2 excess for this object places it more than 4\,$\sigma$ away from the isochrone in the (W1-W2,Z) colour-magnitude diagram in figure 1.   

The diagram in the right panel of figure 6 is for the object noted in sections 4.1.3 and 5.2 above that may be in the process of transition from 
class II to class III.   Its lesser excess at W1 and W2 is clear, as is its relatively greater excess at W3.   Again, the accuracy of the 
original model at these wavelengths is confirmed by the newly observed WISE values.   
While the model shown does preclude the existence of an optically thick inner disk, it does not require the presence of an evacuated 
hole in the inner disk.   Instead, a reduced scale height, i.e. a flatter inner disk is sufficient.   
This evolution to a flatter disk could be caused by grain growth and dust settling along the 
lines proposed by \citet{dul04}.   Ergo, in this model, the process of transition from class II to class III that may be occurring around 
this object's disk does not require the presence of mechanisms (e.g. planet formation) that would completely clear the inner disk.

\begin{figure*}

\begin{tabular}{|c|c}

\includegraphics[scale=0.58]{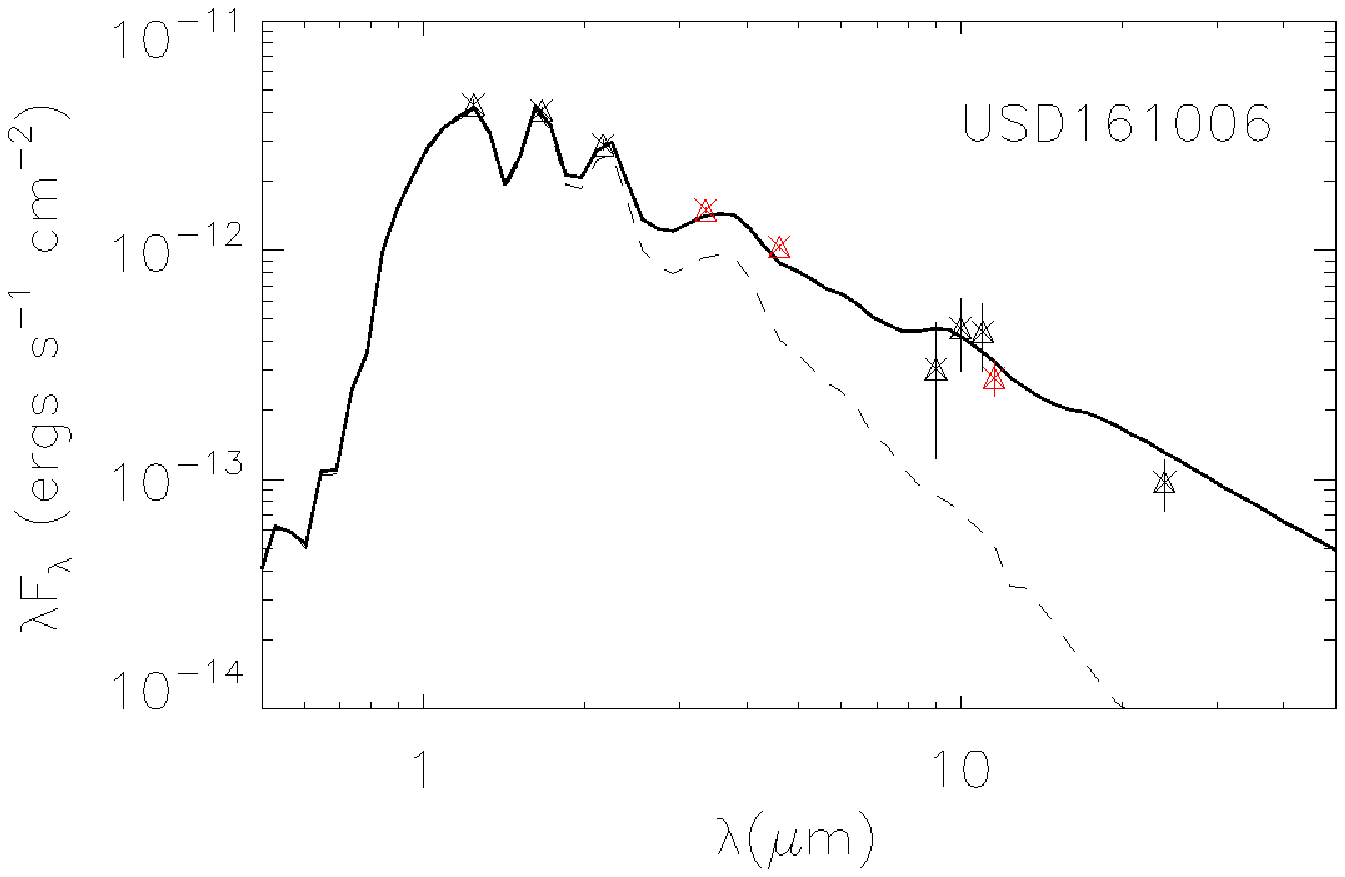}
\includegraphics[scale=0.58]{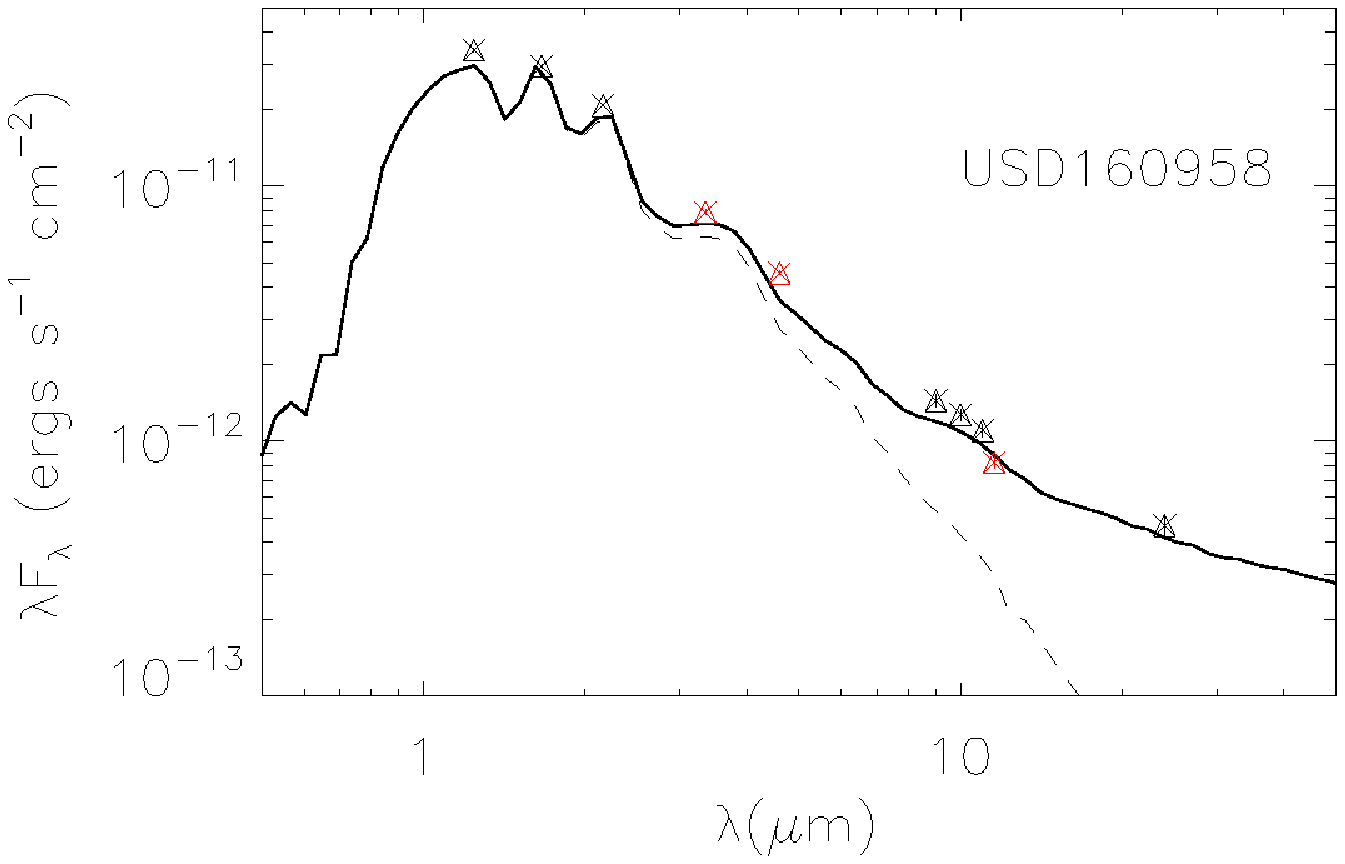}  \\

\end{tabular}
  \caption{Spectral energy distribution for the two class II objects originally identified in \citet{sch07} and recovered in this work.   
Data shown is taken from that of \citet{sch07} for the J, H, K, 9, 10, 11 and 24\,$\mu$m wavelengths, along with the newly acquired W1, W2 
and W3 values.   The Monte Carlo radiative transfer simulations shown for comparison use the same parameters as were given in \citet{sch07}.   
The dotted lines depict calculated photospheric flux while the solid lines represent combined photospheric and disk flux.}
\end{figure*}

\subsection{Comparison with Riaz 2012}

In a recent paper \citet{ria12} analyse a sample of 43 spectroscopically confirmed very low mass members (spectral types M4-M8.5) in UpSco 
using WISE data.   
They find 6 new class II objects and recover 4 others previously recorded by \citet{sch07} and a further 2 that were found by 
\citet{sle08}.   These 12 objects are listed in their Table 1.   We have been kindly provided with a table which includes the 31 
other brown dwarfs that they investigated and categorised as class III by B. Riaz (private communication).   From the 12 class II and 31 class III 
objects that they list, there are 12 in common to both that study and ours.   Both studies categorise the same 5 objects as class II and the 
same 7 objects as class III.   

Also included in the table supplied by B. Riaz are 6 other objects with S/N $<$\,3.0 in the W3 passband.   
\citet{ria12} rely on the use of the W3 signal and require that a source have a W3 S/N of $\geq$\,3.0.   As a result they do not 
categorise these 6 brown dwarfs.   However, using the W1-W2 colour as a primary diagnostic instead of W3 allows 
a larger range of objects to be successfully examined.   All 116 objects investigated in our work have a S/N $>$\,8 in the W1 and W2 passbands, 
while only 64 of them have a S/N of $\geq$\,3.0 in the W3 passband.   The 6 objects which cannot be categorised in \citet{ria12} because 
they have a weak W3 signal have been determined in this work to be class III objects using their W1 and W2 signals alone.   

Requiring that a source have a W3 S/N of $\geq$\,3.0 not only restricts the number of objects in any sample from UpSco, it also produces a sample 
that is biased with respect to the presence of disks.   Class III objects in UpSco are less likely 
than class II objects to have a strong W3 signal.   Of the 116 objects examined in this work, 78\% of the class II objects, but only 48\% of 
the class III objects have a W3 S/N of $\geq$\,3.0.   Restricting the analysis to those 64 objects only would have yielded an artificially high disk 
fraction of 33\%, rather than the 23\% found from the larger unbiased sample of 116 objects, while also increasing the statistical errors.   
For these reasons our disk fraction is a more representative value for the brown dwarf population in UpSco.

\subsection{Comparison with Luhman \& Mamajek 2012}

Another recent paper focused on the disks of UpSco members has been
published by \citet{lam12}. Similar to \citet{ria12}
they look at a sample of spectroscopically confirmed members.
Their list of targets includes 387 objects with spectral types
M4-M8 and another 23 with M8-L2 and is thus substantially larger than
the Riaz et al. sample. They analyse the available Spitzer and WISE
photometry for these objects. To identify the objects with disks, they
use the spectral regime from 4.5 to 24$\,\mu$m, similar to this
paper.

48 of their targets also appear in our total sample of 119, among them 13
classified as class II in this paper. In most cases, our class II/III distinction 
agrees with the one in \citet{lam12}. The only exception is the M9-L1 
object 2MASSJ16082847-2315103, which clearly fulfills our
class II criterion based on its W1-W2 colour, whereas \citet{lam12} 
conclude that it is diskless based on its K-W2 colour. At these
late spectral types, however, the intrinsic colours of young brown dwarfs
increase significantly with spectral type and cover a broad range, i.e.  it is
challenging to unambiguously distinguish between class II and class III. 
Whether this source does or does not have infrared excess emission remains to be 
determined, but this does not affect our results in any way.

The remainder of our sample (71 in total, among them 14 with disks) is
new and not covered by \citet{lam12}, i.e. our work increases
the sample of brown dwarfs analysed with mid-infrared data significantly.

The disk fraction derived by \citet{lam12} for very low mass
and substellar members of UpSco is $\sim 25$\%, and thus agrees very well
with our own result. They also find that disk fractions increase from very
small values for B-G stars ($\le 10$\%) to 25\% for $\ge$\,M5 objects. This
seems at odds with our own statement of spectral type-independent disk
fractions, but a more detailed examination shows that the two studies
actually give consistent results for K and M stars.

\citet{lam12} report disk fractions of 6/67 for K0-M0, 35/231 
for M0-M4, 97/387 for M4-M8 and 4/23 for M8-L2. To compare with our
own values, we calculated the binomial confidence intervals for their
disk fractions and obtain $14\pm 3$\% for K0-M4 and $25\pm 3$\% for M4 to L2.
This is consistent with the numbers quoted in Sect. 5.1. (19\% for
K0-M5 and 23\% for our sample of brown dwarfs). Note that the uncertainties
quoted above only give the statistical confidence interval and do not
take into account possible biases (e.g., age spread across the region, 
uncertainties in spectral types). Thus, based on the current samples, the
evidence for a mass dependence of the disk fraction for objects
later than K0 is marginal at best.

\section{Conclusions}

We have carried out a survey for disks around a homogeneous sample of 119 brown dwarfs, the majority of which have not been previously 
discussed in the literature, in the 5\,Myr old UpSco star forming region 
using photometry from WISE.   Contamination in the sample appears to be negligible and the method of selection is unbiased with respect to
the presence of disks.   

Examining all the UKIDSS, 2MASS and WISE colour-magnitude and colour-colour combinations shows that the WISE W1-W2 
colour is the best primary diagnostic for the presence of a disk around the objects.
   
27 class II objects are identified from the sample.   22 were classified via their W1-W2 colour excess alone.   
5 other objects were also categorised as class II from their W3 and/or W4 signals.   These 5 objects (19\% of all disks) appear to be in the 
transition phase between class II and class III, leading to the conclusion that this phase is short lived, 
lasting less than 0.4\,Myr, an estimate that is consistent with findings for low-mass stars.   

The disk fraction is found to be 23\%$\pm5$\%.   This fraction is statistically 
indistinguishable from results for K/M stars in UpSco.   Results from the literature for Cha I, IC348 and $\sigma$\,Ori show that their brown 
dwarf disk fractions are also indistinguishable from their K/M star disk fractions.   Therefore the average lifetime of the disks in 
each of these regions shows no obvious dependency on the mass of the central object.   Combined with the short transitional phase from 
class II to class III, this suggests that the evolution of brown dwarf disks follows ``a two timescale model'', similar to low mass stars.


\section*{Acknowledgements}

The authors would like to thank Isabelle Baraffe of Exeter University 
and France Allard of the Centre de Recherche Astrophysique de Lyon for 
supplying model data. This work was supported by Science Foundation Ireland
within the Research Frontiers Programme under grant no. 10/RFP/AST2780.
This publication makes use of data products from the Wide-field
Infrared Survey Explorer, which is a joint project of the University of California, Los Angeles,
and the Jet Propulsion Laboratory/California Institute of Technology, funded by 
the National Aeronautics and Space Administration.
This publication also makes use of data products from the Two Micron 
All Sky Survey, which is a joint project of the University of Massachusetts and the 
Infrared Processing and Analysis Center/California Institute of Technology, funded by 
the National Aeronautics and Space Administration and the National Science Foundation.
We would also like to thank the UKIDSS Team for the excellent database they have made 
available to the community.

\begin{table*}
 \begin{minipage}{170mm}
  \caption{Positions, UKIDSS Z and J photometry, WISE W1, W2, W3 and W4 photometry 
of the 27 class II objects.   Objects are listed in order of decreasing mass, based on their Z magnitude. 
Coordinates are J2000.}
  \begin{tabular}{|c|c|c|c|c|c|c|c|c|c}
  \hline
    Name & R.A. & Dec. & Z Mag. & J Mag. & W1 Mag. & W2 Mag. & W3 Mag. & W4 Mag.\\

 \hline
2MASSJ16075049-2125200 & 16:07:50.49 & -21:25:20.2 & 14.03 & 12.69 & 11.27 & 10.73 & 9.10 & 7.22\\
2MASSJ15465432-2556520 & 15:46:54.32 & -25:56:52.1 & 14.16 & 12.75 & 11.40 & 10.81 & 9.10 & 6.23\\
2MASSJ16052875-2655496 & 16:05:28.75 & -26:55:49.7 & 14.18 & 12.89 & 11.67 & 11.21 & 9.90 & 8.03\\
2MASSJ16095852-2345186 & 16:09:58.52 & -23:45:18.7 & 14.20 & 12.57 & 11.34 & 10.97 & 9.96 & 8.73*\\
2MASSJ16030235-2626163 & 16:03:02.36 & -26:26:16.4 & 14.46 & 13.12 & 11.44 & 10.92 & 9.59 & 8.04*\\
2MASSJ15470374-2601183 & 15:47:03.74 & -26:01:18.4 & 15.02 & 13.32 & 12.03 & 11.74 & 10.42 & 7.39\\
2MASSJ16134880-2509006 & 16:13:48.81 & -25:09:00.7 & 15.04 & 13.56 & 11.96 & 11.32 & 9.55 & 7.58\\
2MASSJ15514709-2113234 & 15:51:47.09 & -21:13:23.5 & 15.12 & 13.54 & 10.56 & 9.88 & 8.85 & 7.41\\
2MASSJ16035573-2738248 & 16:03:55.73 & -27:38:25.1 & 15.19 & 13.80 & 12.70 & 12.45 & 12.13* & 7.89\\
2MASSJ15472572-2609185 & 15:47:25.73 & -26:09:18.5 & 15.40 & 13.55 & 11.87 & 11.51 & 9.58 & 7.37\\
2MASSJ16145253-2718557 & 16:14:52.53 & -27:18:55.7 & 15.65 & 14.06 & 12.34 & 11.79 & 10.63 & 8.70*\\
2MASSJ15412655-2613253 & 15:41:26.55 & -26:13:25.4 & 15.65 & 14.00 & 12.59 & 12.10 & 10.75 & 8.24*\\
2MASSJ15521088-2125372 & 15:52:10.88 & -21:25:37.4 & 15.72 & 13.86 & 11.01 & 10.15 & 8.90 & 6.92\\
2MASSJ16143287-2242133 & 16:14:32.87 & -22:42:13.5 & 15.82 & 14.16 & 12.88 & 12.34 & 11.74* & 8.38*\\
2MASSJ15501958-2805237 & 15:50:19.58 & -28:05:23.9 & 16.04 & 14.56 & 13.28 & 12.72 & 12.03* & 8.98*\\
2MASSJ16080745-2345055 & 16:08:07.45 & -23:45:05.6 & 16.06 & 14.40 & 12.86 & 12.21 & 10.30 & 8.43*\\
2MASSJ15524513-2705560 & 15:52:45.13 & -27:05:56.1 & 16.16 & 14.42 & 13.06 & 12.35 & 10.74 & 8.66*\\
2MASSJ15571880-2711567 & 15:57:18.81 & -27:11:56.8 & 16.39 & 14.78 & 13.42 & 12.86 & 11.90* & 9.05*\\
2MASSJ16142144-2339146 & 16:14:21.44 & -23:39:14.8 & 16.59 & 14.97 & 13.57 & 12.99 & 11.95* & 8.25*\\
2MASSJ16012238-2708194 & 16:01:22.39 & -27:08:19.5 & 16.68 & 14.70 & 13.24 & 12.55 & 11.11 & 8.43*\\
2MASSJ16100608-2127440 & 16:10:06.08 & -21:27:44.1 & 16.82 & 14.82 & 13.29 & 12.65 & 11.90* & 8.72*\\
2MASSJ15541998-2135428 & 15:54:19.99 & -21:35:43.0 & 16.85 & 14.98 & 13.26 & 12.57 & 11.22 & 8.39*\\
2MASSJ16083048-2335109 & 16:08:30.49 & -23:35:11.0 & 16.95 & 14.88 & 13.37 & 12.86 & 11.69* & 8.63*\\
2MASSJ16082847-2315103 & 16:08:28.47 & -23:15:10.4 & 17.64 & 15.45 & 13.77 & 13.22 & 12.21* & 8.34*\\
2MASSJ15472282-2139141 & 15:47:22.82 & -21:39:14.3 & 17.97 & 15.65 & 13.69 & 13.07 & 11.63* & 8.67*\\
2MASSJ15433947-2535549 & 15:43:39.47 & -25:35:54.9 & 18.15 & 15.77 & 13.91 & 13.73 & 10.77 & 7.78\\
2MASSJ15553614-2546591 & 15:55:36.15 & -25:46:59.2 & 18.23 & 15.83 & 13.63 & 13.12 & 11.10* & 7.92\\

\hline
 & & & & & & & &*S/N $<$\,5.

\end{tabular}
\end{minipage}
\end{table*}

\begin{table*}
 \begin{minipage}{170mm}
  \caption{Positions, UKIDSS Z and J photometry, WISE W1, W2, W3 and W4 photometry 
of the 89 class III objects.   Objects are listed in order of decreasing mass, based on their Z magnitude. 
Coordinates are J2000.}
  \begin{tabular}{|c|c|c|c|c|c|c|c|c|c}
  \hline
    Name & R.A. & Dec. & Z Mag. & J Mag. & W1 Mag. & W2 Mag. & W3 Mag. & W4 Mag.\\

\hline

2MASSJ16175608-2856399 & 16:17:56.09 & -28:56:40.0 & 14.01 & 12.76 & 11.76 & 11.53 & 10.54 & 8.17*\\
2MASSJ16034797-2801319 & 16:03:47.97 & -28:01:31.9 & 14.03 & 12.81 & 11.83 & 11.62 & 11.17 & 9.02*\\
2MASSJ16105728-2359540 & 16:10:57.28 & -23:59:54.1 & 14.04 & 12.80 & 11.68 & 11.48 & 11.72* & 8.72*\\
2MASSJ15554229-2546477 & 15:55:42.29 & -25:46:47.8 & 14.07 & 12.65 & 11.52 & 11.24 & 10.93* & 8.18*\\
2MASSJ16055898-2556228 & 16:05:58.99 & -25:56:22.9 & 14.10 & 12.59 & 11.50 & 11.22 & 10.83 & 9.08*\\
2MASSJ16105429-2309108 & 16:10:54.29 & -23:09:11.1 & 14.13 & 12.97 & 11.88 & 11.68 & 11.60* & 8.27*\\
2MASSJ15591513-2840411 & 15:59:15.12 & -28:40:41.3 & 14.14 & 12.96 & 12.00 & 11.77 & 12.39* & 8.75*\\
2MASSJ16095217-2136277 & 16:09:52.17 & -21:36:27.8 & 14.15 & 12.51 & 11.33 & 10.97 & 10.72 & 9.14*\\
2MASSJ16063691-2720548 & 16:06:36.91 & -27:20:54.9 & 14.16 & 12.88 & 11.85 & 11.60 & 11.91* & 8.73*\\
2MASSJ15411513-2539447 & 15:41:15.14 & -25:39:44.8 & 14.16 & 12.76 & 11.59 & 11.37 & 11.30 & 8.71*\\
2MASSJ16033799-2611544 & 16:03:37.99 & -26:11:54.4 & 14.17 & 12.98 & 11.96 & 11.70 & 11.22 & 8.63*\\
2MASSJ16105499-2126139 & 16:10:54.99 & -21:26:14.0 & 14.22 & 12.73 & 11.56 & 11.30 & 10.89 & 8.53*\\
2MASSJ16154869-2710546 & 16:15:48.69 & -27:10:54.7 & 14.22 & 12.82 & 11.69 & 11.40 & 11.65* & 8.86*\\
2MASSJ15450519-2559047 & 15:45:05.20 & -25:59:04.7 & 14.23 & 12.90 & 11.76 & 11.54 & 12.00* & 8.53*\\
2MASSJ16072196-2358452 & 16:07:21.96 & -23:58:45.3 & 14.24 & 12.99 & 11.86 & 11.64 & 11.27 & 8.79*\\
2MASSJ16152819-2315439 & 16:15:28.19 & -23:15:44.1 & 14.24 & 13.12 & 12.08 & 11.90 & 11.78* & 8.18*\\
2MASSJ16370523-2625439 & 16:37:05.24 & -26:25:44.0 & 14.27 & 12.96 & 11.96 & 11.69 & 11.40 & 8.91*\\
2MASSJ15492909-2815384 & 15:49:29.08 & -28:15:38.6 & 14.29 & 12.96 & 11.85 & 11.63 & 12.05* & 8.91*\\
2MASSJ15491602-2547146 & 15:49:16.02 & -25:47:14.6 & 14.31 & 13.01 & 11.87 & 11.60 & 10.90 & 8.20*\\
2MASSJ16082229-2217029 & 16:08:22.29 & -22:17:03.0 & 14.31 & 12.94 & 11.85 & 11.56 & 11.10 & 8.81*\\
2MASSJ16061595-2218279 & 16:06:15.95 & -22:18:28.0 & 14.31 & 13.17 & 12.11 & 11.91 & 11.92* & 9.02*\\
2MASSJ15544260-2626270 & 15:54:42.61 & -26:26:27.0 & 14.32 & 13.05 & 11.91 & 11.72 & 11.61* & 8.23*\\
2MASSJ16090168-2740521 & 16:09:01.68 & -27:40:52.3 & 14.33 & 12.86 & 11.71 & 11.44 & 10.97 & 8.72*\\
2MASSJ15582376-2721435 & 15:58:23.76 & -27:21:43.7 & 14.35 & 13.07 & 12.02 & 11.79 & 11.37* & 8.78*\\
2MASSJ16132180-2731219 & 16:13:21.80 & -27:31:22.0 & 14.36 & 13.25 & 11.93 & 11.85 & 11.60* & 8.77*\\
2MASSJ15415562-2538465 & 15:41:55.63 & -25:38:46.5 & 14.37 & 13.10 & 12.02 & 11.79 & 12.14* & 8.93*\\
2MASSJ16002535-2644060 & 16:00:25.35 & -26:44:06.1 & 14.38 & 13.02 & 11.90 & 11.66 & 11.67* & 8.50*\\
2MASSJ15545410-2114526 & 15:54:54.11 & -21:14:52.7 & 14.40 & 13.26 & 12.10 & 11.94 & 10.71 & 8.31*\\
2MASSJ16062637-2306113 & 16:06:26.37 & -23:06:11.4 & 14.48 & 13.20 & 12.12 & 11.87 & 11.68* & 8.90*\\
2MASSJ16121609-2344248 & 16:12:16.09 & -23:44:25.0 & 14.50 & 13.19 & 11.96 & 11.78 & 10.99 & 8.21*\\
2MASSJ16090451-2224523 & 16:09:04.51 & -22:24:52.5 & 14.57 & 12.92 & 11.66 & 11.40 & 10.59 & 8.23*\\
2MASSJ16113470-2219442 & 16:11:34.70 & -22:19:44.3 & 14.61 & 13.24 & 12.11 & 11.87 & 12.25* & 8.77*\\
2MASSJ15505993-2537116 & 15:50:59.94 & -25:37:11.7 & 14.62 & 13.33 & 12.30 & 12.05 & 11.55* & 8.90*\\
2MASSJ15524857-2621453 & 15:52:48.57 & -26:21:45.4 & 14.62 & 13.30 & 12.27 & 12.00 & 12.51* & 9.14*\\
2MASSJ16372782-2641406 & 16:37:27.83 & -26:41:40.7 & 14.63 & 13.30 & 12.14 & 12.00 & 11.42 & 8.95*\\
2MASSJ15522943-2721003 & 15:52:29.44 & -27:21:00.4 & 14.64 & 13.47 & 12.45 & 12.14 & 12.36* & 9.06*\\
2MASSJ15530374-2600306 & 15:53:03.75 & -26:00:30.7 & 14.66 & 13.43 & 12.36 & 12.14 & 12.03* & 8.62*\\
2MASSJ15493660-2815141 & 15:49:36.59 & -28:15:14.3 & 14.66 & 13.39 & 12.36 & 12.05 & 11.63* & 8.79*\\
2MASSJ16133476-2328156 & 16:13:34.76 & -23:28:15.7 & 14.74 & 13.48 & 12.39 & 12.12 & 11.77* & 8.81*\\
2MASSJ15490803-2839550 & 15:49:08.02 & -28:39:55.2 & 14.82 & 13.60 & 12.45 & 12.21 & 10.68 & 8.15*\\
2MASSJ16112630-2340059 & 16:11:26.30 & -23:40:06.1 & 14.83 & 13.40 & 12.16 & 11.92 & 12.04* & 8.55*\\
2MASSJ15495733-2201256 & 15:49:57.33 & -22:01:25.7 & 14.89 & 13.35 & 12.15 & 11.89 & 11.56* & 8.64*\\
2MASSJ16062870-2856580 & 16:06:28.70 & -28:56:58.2 & 14.90 & 13.52 & 12.39 & 12.18 & 11.41 & 9.06*\\
2MASSJ15572692-2715094 & 15:57:26.93 & -27:15:09.5 & 14.93 & 13.66 & 12.59 & 12.39 & 12.00* & 9.10*\\
2MASSJ16164539-2333413 & 16:16:45.39 & -23:33:41.6 & 14.99 & 13.73 & 12.62 & 12.33 & 11.61* & 8.44*\\
2MASSJ16005265-2812087 & 16:00:52.66 & -28:12:09.0 & 15.04 & 13.57 & 12.48 & 12.27 & 11.79* & 9.13*\\
2MASSJ16132665-2230348 & 16:13:26.66 & -22:30:35.0 & 15.05 & 13.52 & 12.33 & 11.99 & 11.00 & 8.82*\\
2MASSJ16115737-2215066 & 16:11:57.37 & -22:15:06.8 & 15.07 & 13.67 & 12.51 & 12.24 & 11.76* & 8.84*\\
2MASSJ16064910-2216382 & 16:06:49.10 & -22:16:38.4 & 15.09 & 13.73 & 12.63 & 12.37 & 11.90* & 8.79*\\
2MASSJ15585793-2758083 & 15:58:57.93 & -27:58:08.5 & 15.13 & 13.81 & 12.78 & 12.54 & 12.08* & 9.06*\\
2MASSJ15420830-2621138 & 15:42:08.31 & -26:21:13.8 & 15.15 & 13.74 & 12.59 & 12.32 & 12.10* & 9.07*\\
2MASSJ16090197-2151225 & 16:09:01.98 & -21:51:22.7 & 15.16 & 13.59 & 12.32 & 12.08 & 10.87 & 8.45*\\
2MASSJ16124692-2338408 & 16:12:46.92 & -23:38:40.9 & 15.18 & 13.60 & 12.42 & 12.19 & 12.20* & 8.25*\\
2MASSJ16153648-2315175 & 16:15:36.48 & -23:15:17.6 & 15.19 & 13.93 & 12.85 & 12.61 & 12.03* & 8.23*\\
2MASSJ16134264-2301279 & 16:13:42.64 & -23:01:28.0 & 15.19 & 13.72 & 12.47 & 12.25 & 12.20* & 8.38*\\
2MASSJ16113837-2307072 & 16:11:38.37 & -23:07:07.5 & 15.19 & 13.74 & 12.60 & 12.31 & 11.97* & 8.59*\\
2MASSJ15583403-2803243 & 15:58:34.03 & -28:03:24.5 & 15.21 & 13.72 & 12.53 & 12.30 & 11.45 & 9.03*\\
2MASSJ16192399-2818374 & 16:19:23.99 & -28:18:37.5 & 15.29 & 13.79 & 12.73 & 12.40 & 11.51* & 8.57*\\
2MASSJ15490414-2120150 & 15:49:04.14 & -21:20:15.2 & 15.31 & 13.77 & 12.64 & 12.37 & 11.50* & 8.92*\\
2MASSJ16051243-2624513 & 16:05:12.43 & -26:24:51.4 & 15.40 & 14.06 & 12.93 & 12.76 & 12.29* & 9.08*\\
2MASSJ15533067-2617307 & 15:53:30.68 & -26:17:30.7 & 15.49 & 13.99 & 12.83 & 12.60 & 12.06* & 8.64*\\
\end{tabular}
\end{minipage}
\end{table*}

\begin{table*}
 \begin{minipage}{170mm}
  \caption{Table 3 continued.}
  \begin{tabular}{|c|c|c|c|c|c|c|c|c|c}
  \hline
    Name & R.A. & Dec. & Z Mag. & J Mag. & W1 Mag. & W2 Mag. & W3 Mag. & W4 Mag\\

\hline

2MASSJ15544486-2843078 & 15:54:44.85 & -28:43:07.9 & 15.51 & 14.12 & 12.99 & 12.81 & 12.60* & 8.90*\\
2MASSJ15531698-2756369 & 15:53:16.98 & -27:56:37.2 & 15.53 & 13.96 & 12.84 & 12.55 & 12.50* & 8.58*\\
2MASSJ15551960-2751207 & 15:55:19.59 & -27:51:21.0 & 15.60 & 14.03 & 12.93 & 12.67 & 12.30* & 9.10*\\
2MASSJ15564227-2646467 & 15:56:42.28 & -26:46:46.8 & 15.62 & 14.03 & 12.85 & 12.54 & 11.65* & 8.51*\\
2MASSJ16101316-2856308 & 16:10:13.15 & -28:56:31.0 & 15.67 & 14.06 & 12.89 & 12.59 & 12.16* & 9.09*\\
2MASSJ16115439-2236491 & 16:11:54.39 & -22:36:49.3 & 15.73 & 14.27 & 13.04 & 12.80 & 11.86* & 8.32*\\
2MASSJ16092938-2343121 & 16:09:29.39 & -23:43:12.2 & 15.96 & 14.20 & 12.94 & 12.65 & 12.30* & 8.98*\\
2MASSJ16103014-2315167 & 16:10:30.14 & -23:15:16.8 & 16.06 & 14.37 & 13.05 & 12.75 & 11.97* & 8.35*\\
2MASSJ15561721-2638171 & 15:56:17.21 & -26:38:17.2 & 16.10 & 14.52 & 13.26 & 13.00 & 10.38 & 8.23*\\
2MASSJ16072641-2144169 & 16:07:26.41 & -21:44:17.1 & 16.17 & 14.60 & 13.43 & 13.17 & 12.26* & 8.55*\\
2MASSJ16142061-2745497 & 16:14:20.61 & -27:45:49.8 & 16.31 & 14.49 & 13.22 & 12.93 & 12.26* & 8.69*\\
2MASSJ15572820-2708430 & 15:57:28.21 & -27:08:43.0 & 16.40 & 14.78 & 13.71 & 13.47 & 12.21* & 8.77*\\
2MASSJ16134079-2219459 & 16:13:40.79 & -22:19:46.1 & 16.49 & 14.72 & 13.42 & 13.08 & 12.05* & 8.99*\\
2MASSJ15442275-2136092 & 15:44:22.75 & -21:36:09.3 & 16.55 & 15.01 & 13.69 & 13.48 & 11.68* & 8.55*\\
2MASSJ15543065-2536054 & 15:54:30.65 & -25:36:05.5 & 16.73 & 15.03 & 13.80 & 13.61 & 11.58* & 8.17*\\
2MASSJ16064818-2230400 & 16:06:48.18 & -22:30:40.1 & 16.82 & 14.93 & 13.63 & 13.37 & 12.04* & 8.68*\\
2MASSJ15444172-2619052 & 15:44:41.72 & -26:19:05.3 & 17.16 & 15.21 & 13.86 & 13.63 & 12.26* & 8.86*\\
2MASSJ16072382-2211018 & 16:07:23.82 & -22:11:02.0 & 17.24 & 15.20 & 13.70 & 13.41 & 12.20* & 8.87*\\
2MASSJ16104714-2239492 & 16:10:47.13 & -22:39:49.4 & 17.44 & 15.26 & 13.80 & 13.51 & 12.02* & 8.57*\\
2MASSJ16084744-2235477 & 16:08:47.44 & -22:35:47.9 & 17.74 & 15.69 & 14.30 & 14.11 & 12.28* & 8.79*\\
2MASSJ15491331-2614075 & 15:49:13.32 & -26:14:07.5 & 18.29 & 16.04 & 14.50 & 14.12 & 12.16* & 8.88*\\
2MASSJ16081843-2232248 & 16:08:18.43 & -22:32:25.0 & 18.51 & 16.10 & 14.28 & 14.05 & 11.56* & 8.72*\\
2MASSJ16195827-2832276 & 16:19:58.26 & -28:32:27.8 & 18.74 & 16.18 & 14.42 & 14.10 & 12.47* & 8.77*\\
2MASSJ15451990-2616529 & 15:45:19.91 & -26:16:53.0 & 18.77 & 16.27 & 14.41 & 14.06 & 11.85* & 8.20*\\
2MASSJ16362646-2720024 & 16:36:26.47 & -27:20:02.5 & 18.77 & 16.33 & 14.52 & 14.22 & 11.44* & 8.79*\\
2MASSJ16360175-2703305 & 16:36:01.75 & -27:03:30.5 & 19.12 & 16.64 & 14.95 & 14.91 & 12.07* & 8.86*\\
2MASSJ16073799-2242468 & 16:07:37.99 & -22:42:47.0 & 19.24 & 16.76 & 15.16 & 14.73 & 12.51* & 8.25*\\
2MASSJ15504498-2554213 & 15:50:44.99 & -25:54:21.4 & 19.90 & 16.93 & 14.86 & 14.70 & 11.84* & 8.33*\\

\hline
 & & & & & & & &*S/N $<$\,5.
\end{tabular}
\end{minipage}
\end{table*}

\appendix

\section[]{New Objects From UKIDSS 9th Data Release}
This appendix outlines the method used to identify new brown dwarfs in UpSco using the UKIDSS 9th Data Release.   
It is the same method used by \citet{dsr11} to analyse the 8th Data Release.   For the sake of brevity, some of the details 
described by \citet{dsr11} are not repeated here.



\begin{figure*}

\begin{tabular}{|c|c}

\includegraphics[scale=0.64]{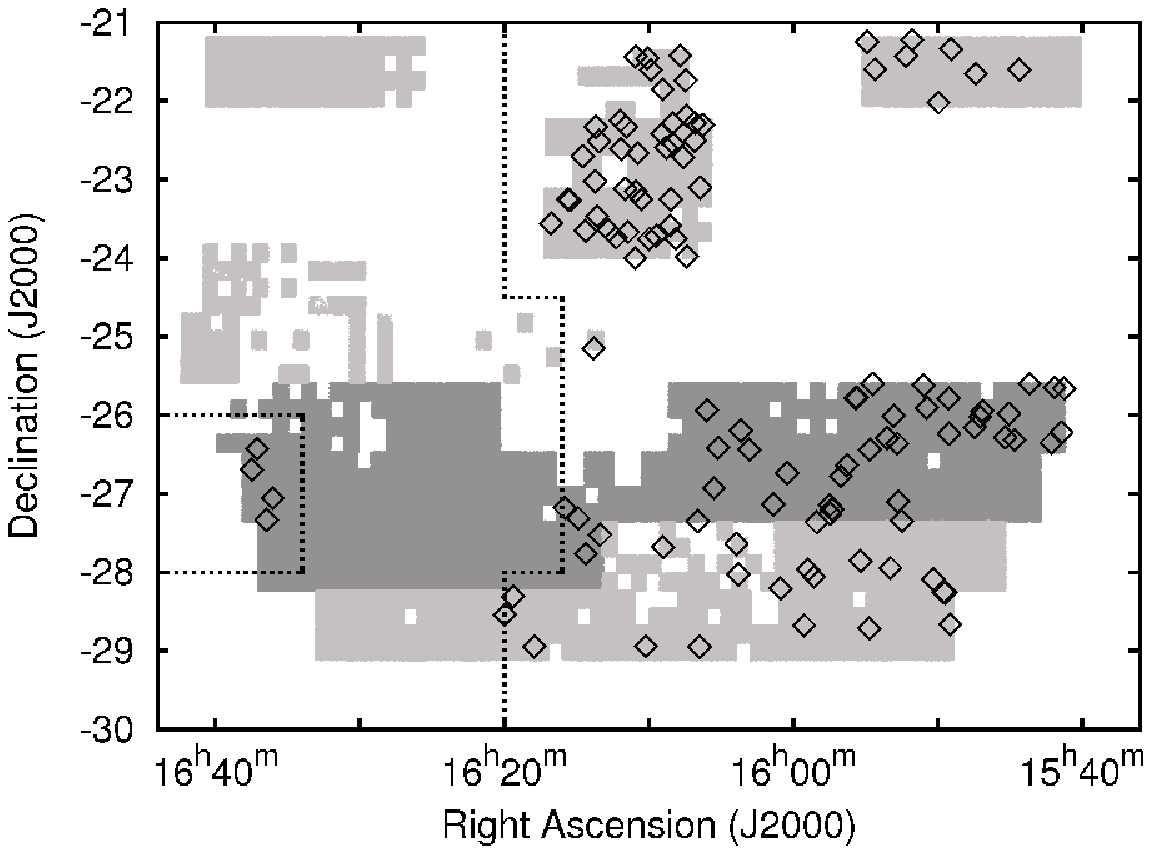}
\includegraphics[scale=0.64]{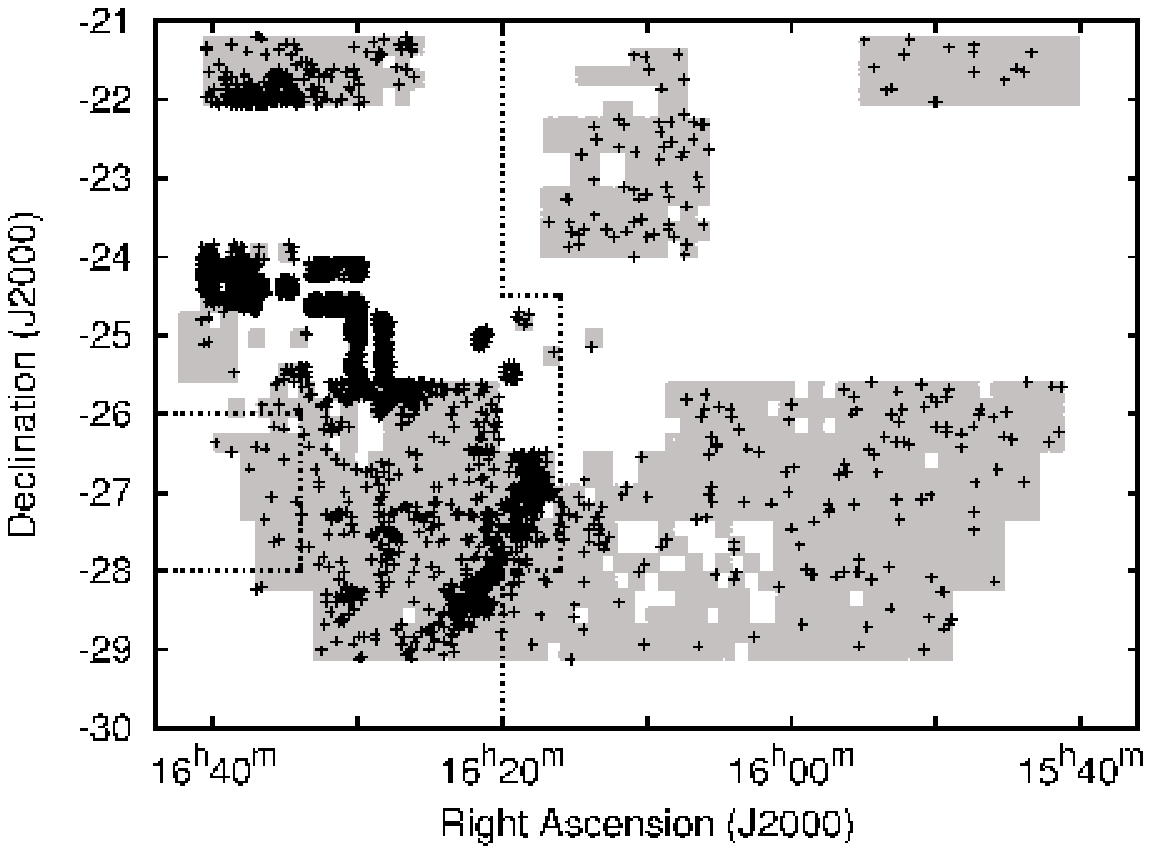}  \\

\end{tabular}
\caption{Coverage in Z, Y, J, H and K filters of 57\,deg$^2$ in Upper Scorpius from the UKIDSS GCS.   The left panel shows 
the new 24\,deg$^2$ of coverage from the 9th Data Release in dark grey and earlier coverage in light grey.   The open diamonds 
mark the position of the 116 brown dwarfs that comprise the sample analysed in this work.   The right panel shows the location of 
objects selected by the first cut in the (Z-J,Z) colour-magnitude diagram.   95\% of all the objects selected from the 9th Data Release are 
clustered around the heavily extincted region surrounding $\rho$\,Oph.   To obtain a minimally contaminated sample of brown dwarf candidates, 
the area inside the two sets of dashed lines was excluded from consideration.}
\end{figure*}

UKIDSS is made up of several components including the 
Galactic Cluster Survey (GCS). Described in detail in \citet{lawrence07} the 
GCS is a survey of ten large open star clusters and star forming regions, including 
UpSco. 

The instrument used to take the GCS images is the Wide Field 
Camera (WFCAM).  Data collected by the 
WFCAM is subject to an automated process that detects and parameterises objects 
and performs photometric and astrometric calibrations.  The resulting reduced 
image frames and catalogues are then placed in the WFCAM Science Archive (WSA).
The WSA can be interrogated using Structured Query Language (SQL).


As shown in figure A1, the new area in 
UpSco investigated here and surveyed for the 9th Data Release covers 24\,deg$^2$.   
The data for objects in the target area were obtained via an SQL query 
to the UKIDSS 
GCS database.   All queries were structured to include only point source 
objects in order to avoid contamination by extended sources (e.g. relatively 
nearby galaxies).   
As every object with photometric 
characteristics consistent with a brown dwarf had its proper motion assessed, 
in order to check whether it is likely a member of UpSco, each 
query submitted also correlated all objects found in the UKIRT GCS databases 
with those found in 2MASS databases. 
The 2MASS data is 
used as a first epoch for the purposes of proper motion calculation.


\subsection{Photometry}





\begin{figure}
  \includegraphics[width=0.45\textwidth]{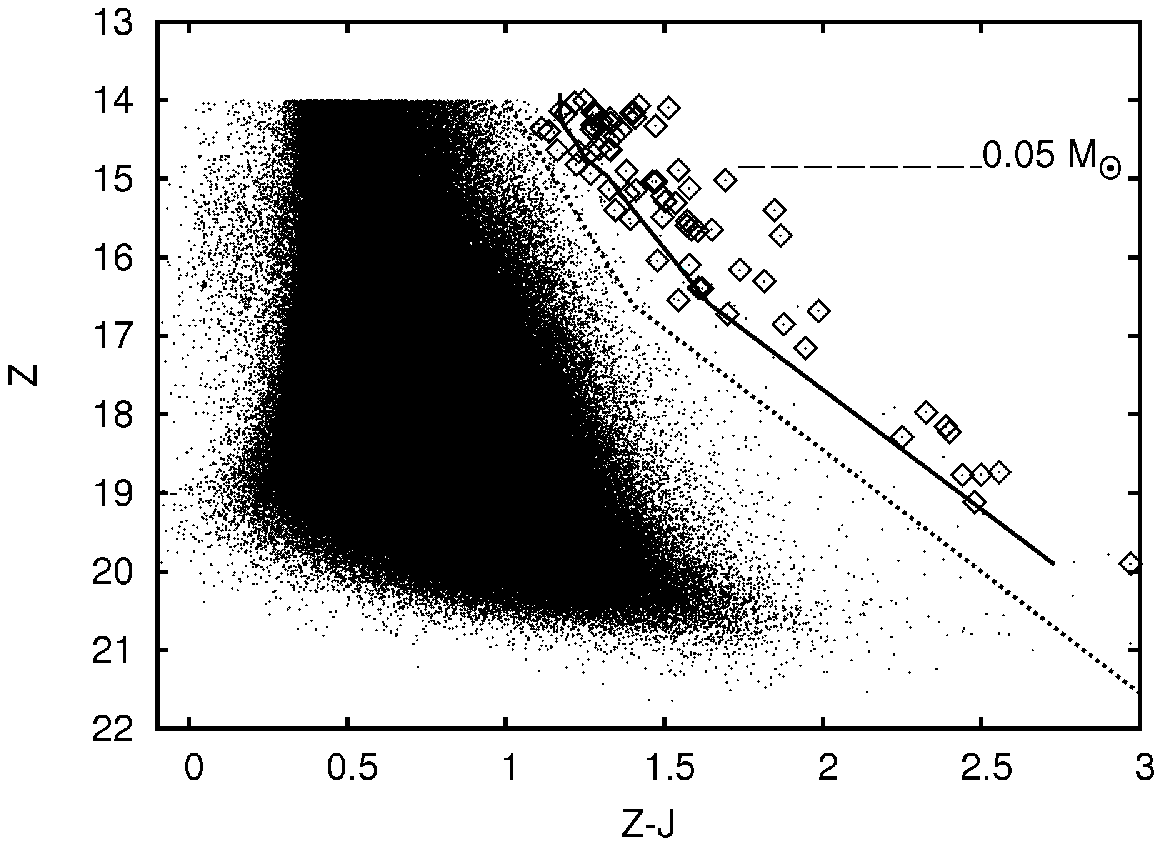}
  \caption{(Z-J,Z) colour-magnitude diagram showing the 76 brown dwarf candidates from the 9th Data Release as open diamonds.   All other objects 
are shown as small dots.   The 5 Myr DUSTY model \citep{cbah00} isochrone 
is also shown with mass decreasing going down the isochrone from 0.09$\,M_{\sun}$ at the top to 0.01$\,M_{\sun}$ at the bottom.   
The 0.05$\,M_{\sun}$ position on the isochrone is indicated.   All objects to the left of the dashed line were rejected because of their colours.}
\end{figure}

A query similar to that shown in \citet{dsr11} was submitted to the WSA. The query returned 
1,438,887 objects.   
The objects were assessed on the basis of their position on a (Z-J,Z) colour-magnitude diagram as shown in figure A2.   
To refine the search, a 
new query was submitted to the WSA eliminating all objects to the left of a 
line in the (Z-J,Z) colour-magnitude diagram from (Z-J,Z) = (1.0, 14.0) through (1.4, 16.6) to (3.0, 21.55) (dashed line in figure A2). 
This query left 4,398 objects.   Reddening caused by extinction shifts objects to the right and down on colour-magnitude diagrams. 
To assess if reddening was contaminating the results the 4,398 objects had their location plotted as shown in the right panel of figure A1.   
There is an obvious clustering of objects in a large area which coincides with the heavily extincted region around $\rho$\,Oph.   
Therefore, the analysis was confined to 706,638 objects in the 9th Data Release that were outside that region (dashed lines in figure A1).   
This left only 200 of the 4,398 objects selected in the initial (Z-J,Z) cut.   
These 200 objects were examined again in the (Z-J,Z) colour-magnitude diagram.   86 of the objects to 
the left of the line (Z-J,Z) = (1.1, 14.0) through (1.1, 14.3), (1.2, 14.9), (1.3, 15.2), (1.6, 17.0) to (3.0,21.0) 
were rejected for being too far from the isochrone on the blue side, leaving 114 photometric candidates.   

\subsection{Proper Motion}

\begin{figure}
  \includegraphics[width=0.45\textwidth]{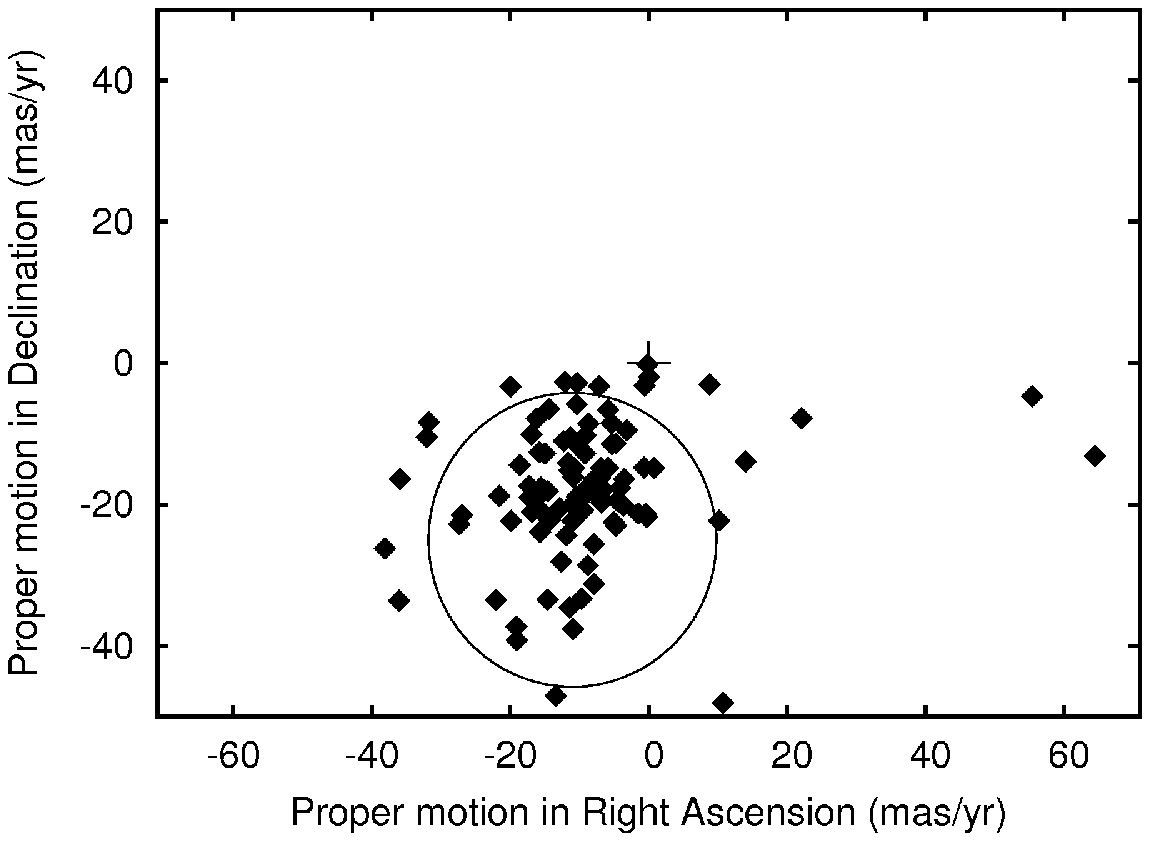}
  \caption{Vector point diagram for 96 candidate brown dwarfs in Upper Scorpius.   There is an obvious and identifiable cluster 
around (-11,-25), while there is no significant clustering around the origin, indicating that there is very little contamination 
from background objects in the sample.   Candidates lying outside the 2\,$\sigma$ selection circle were classified as non-members, 
leaving 76 brown dwarf members of UpSco.}
\end{figure}

The 114 photometric candidates were then examined to find their 
proper motion.  
The resulting vector point diagram is shown in figure A3.   The known 
proper motions of UpSco in right ascension and declination are about -11mas/yr and -25mas/yr 
respectively \citep{deb97,pre02}.   Of the 114 candidates, 4 were too faint to be recorded in 2MASS 
leaving 110 candidates with proper motion data calculated.  
The remaining 110 candidates included 14 with proper motions greater than 
the range of figure A3.   


All 96 candidates shown in figure A3 are predominantly centred around the (-11,-25) position.   
A 2\,$\sigma$ selection circle as calculated in \citet{dsr11} is shown centred on that position.   
There is no significant clustering of objects around the (0,0) position indicating that the sample is not contaminated by more distant 
objects e.g. AGB stars which have similar surface temperatures and colours to 
brown dwarfs, but much greater intrinsic luminosities.   The 76 candidates within the 2\,$\sigma$ selection circle 
were then classified as members of UpSco.   These objects so selected 
have the photometric and proper motion characteristics of a 5~Myr old 
brown dwarf member of UpSco.   
25 of the 76 were among the sample of 68 selected by \citet{dsr11}, yielding 
51 new objects from the 9th Data Release to add to the sample.   
From these 51 brown dwarfs, 5 have been identified before.  
\citet{lodieu06} identifies 1 of them, \citet{mar04} lists another and \citet{ard00} identifies 3.   
We note that the remaining 46 have not been previously identified in other surveys.

One potential caveat with respect to this sample is the fact that the Upper Scorpius
population and the younger $\rho$\,Oph population are unlikely to be clearly separated in colour-magnitude or proper motion 
diagrams.   By avoiding the areas with very high extinction (see figure A1) we exclude the bulk of the population of $\rho$\,Oph, but some
scattered young substellar members of $\rho$\,Oph might still contaminate our sample.   The low disk fraction in our sample, however,
suggests that this cannot be a major factor.

\section[]{Selected Spectral Energy Distributions}
In this appendix we show spectral energy distributions for 12 objects from our sample, 6 class II and 6 class III objects for comparison.   
The class III objects used have the closest magnitude in the J passband to the class II objects with which they are compared.   
Shown in the top panels of figure B1 are 2 of the typical examples that were deemed class II on the basis of their W1-W2 (3.4\,$\mu$m-4.6\,$\mu$m) 
colour alone (as is the object featured in the left panel in figure 6).   
The difference in slopes between the W1 and W2 points on the class II and class III spectral energy distributions are 
apparent, as are the higher W3 (12\,$\mu$m) and W4 (22\,$\mu$m) values of the class II objects.   
The 2 class II objects that stand out in figure 4 because of their large J-K excess are shown in the middle panels of figure B1.   
Of all the 27 class II objects, the spectral 
energy distributions for these 2 show the greatest divergence from those of their corresponding class III obects .   
The bottom panels in figure B1 show 2 of the 5 class II objects that could not be distinguished on the basis of their W1-W2 colour alone.   
(The right panel in figure 6 features another of these objects).   
The bottom right panel is the most extreme example of all the 5 objects with little or no W1-W2 excess and deemed class II based 
on their bright W3 and/or W4 signals (the transition disks).   
It exhibits no W1-W2 excess and has a weak W3 signal.   Apart from its strong W4 signal it resembles a class III object.

\begin{figure*}

\begin{tabular}{|c|c}

\includegraphics[scale=0.58]{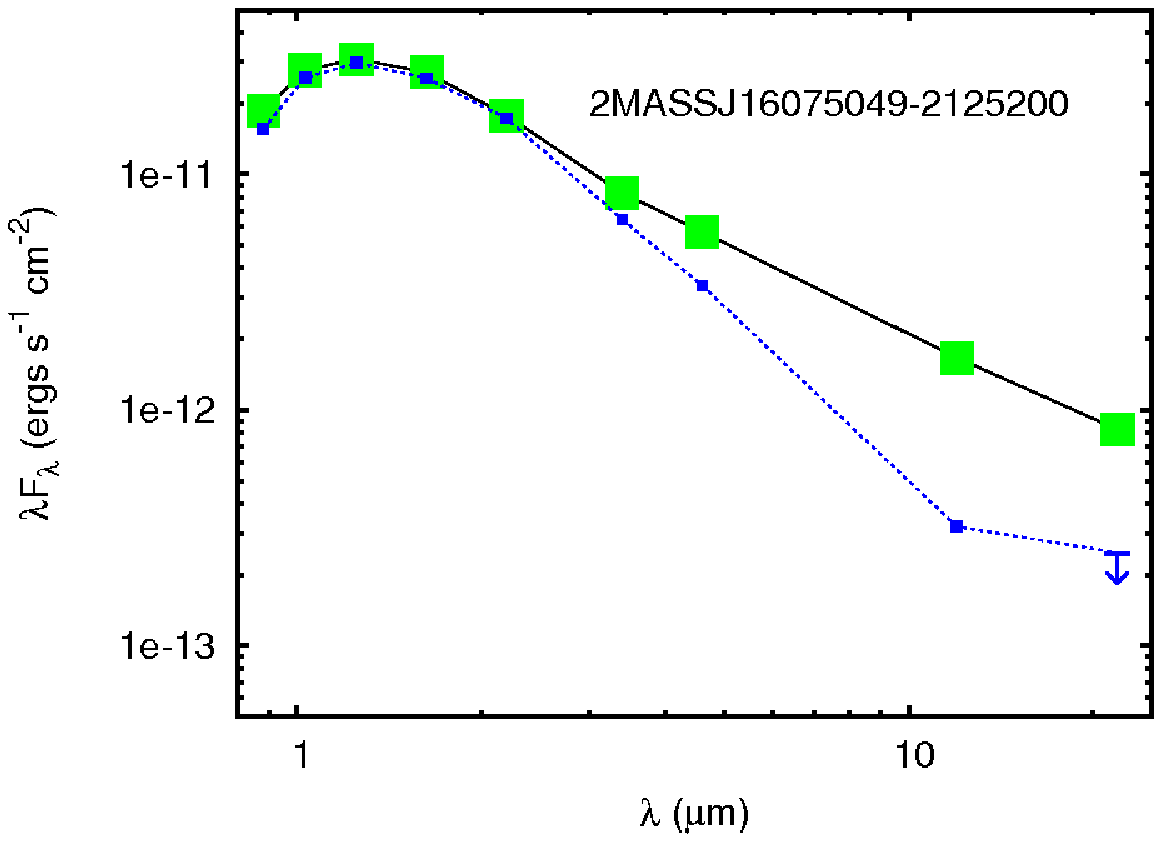}
\includegraphics[scale=0.58]{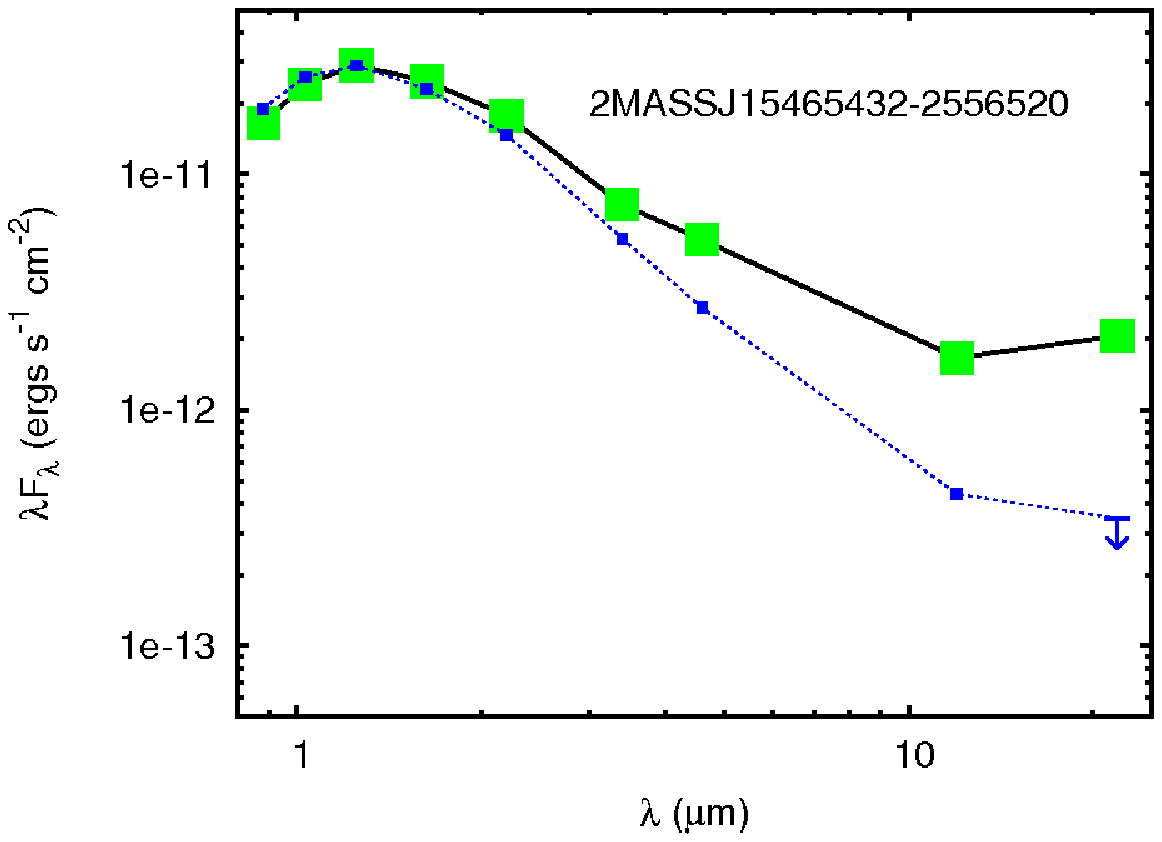}  \\
\includegraphics[scale=0.58]{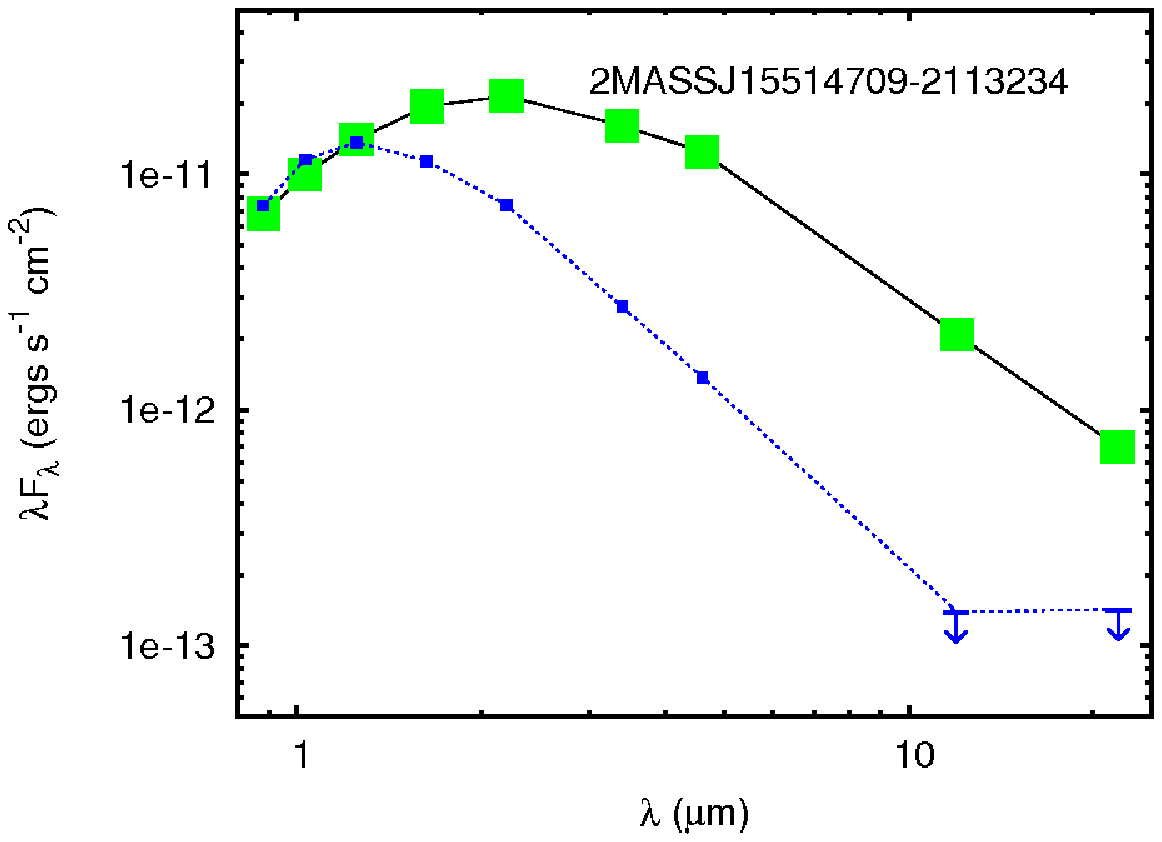}
\includegraphics[scale=0.58]{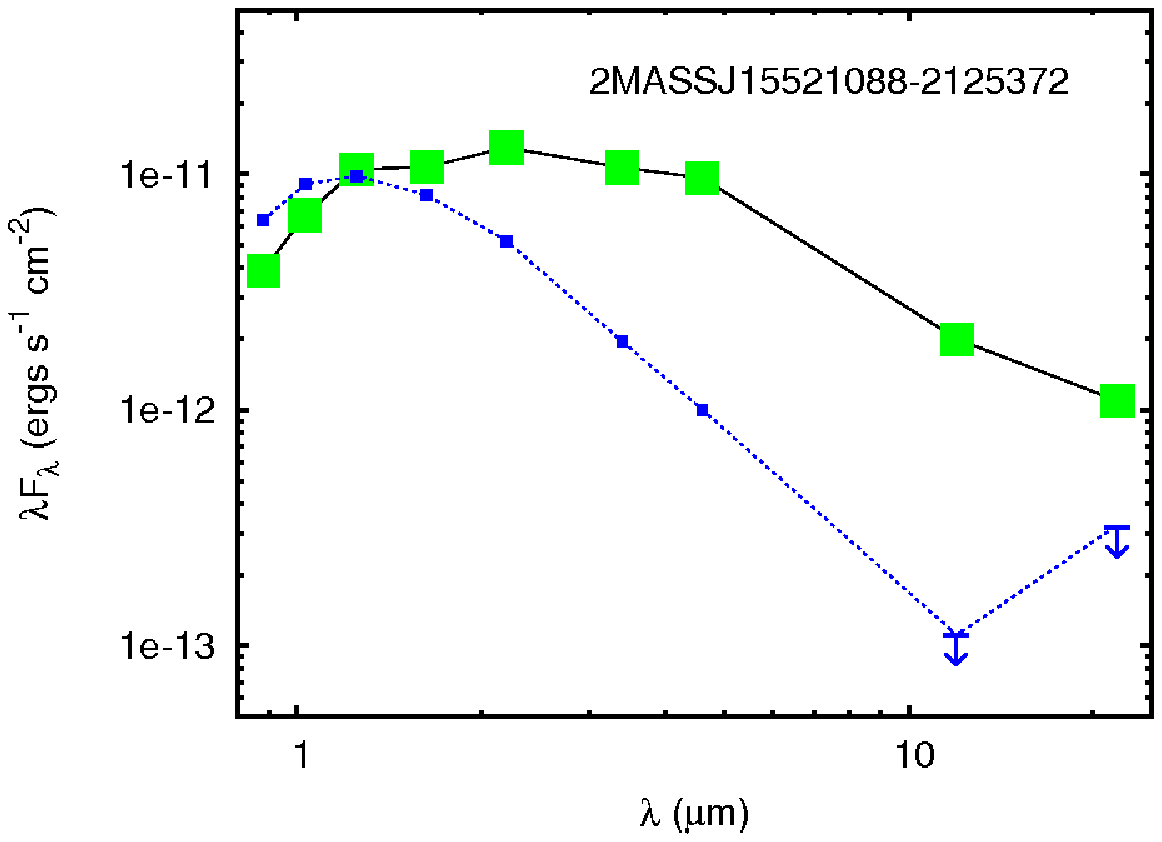}  \\
\includegraphics[scale=0.58]{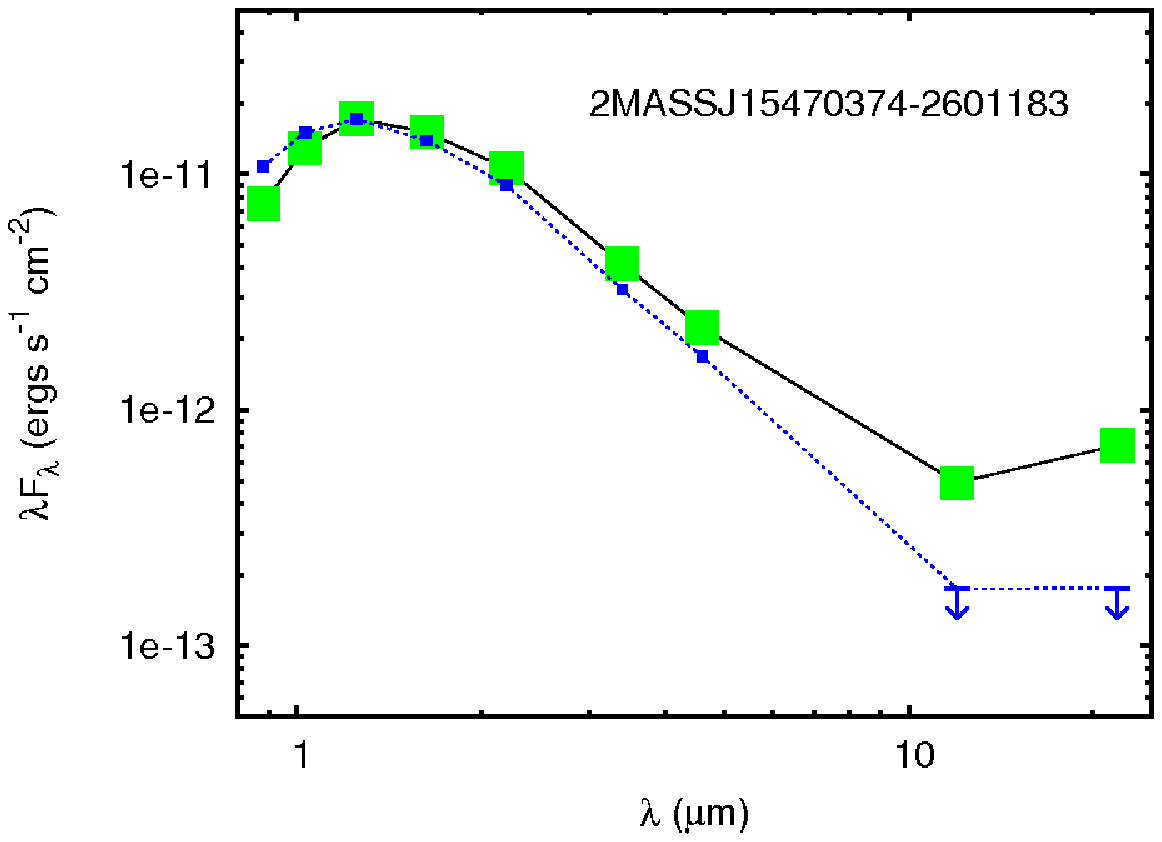}
\includegraphics[scale=0.58]{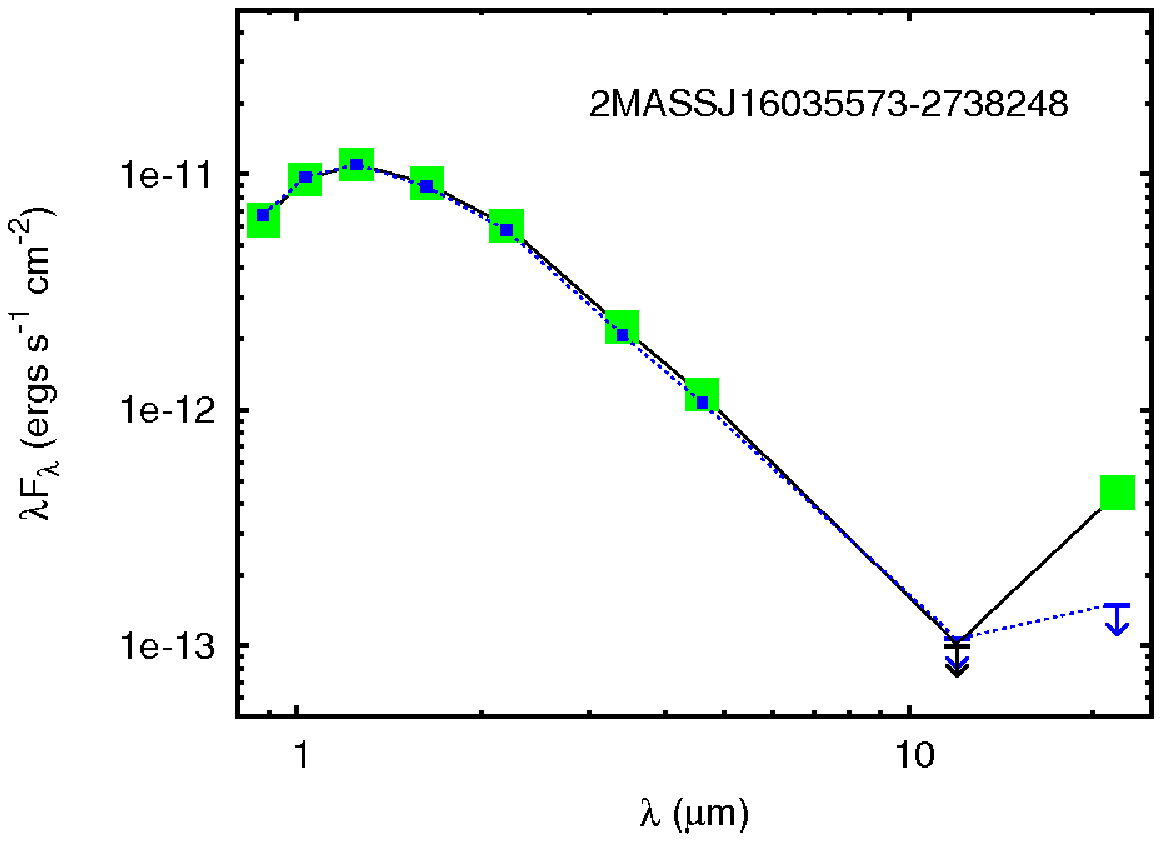}  \\

\end{tabular}
  \caption{Spectral energy distributions for 6 of the 27 class II objects (solid lines with large squares) shown alongside 6 of the 89 class III 
objects (dotted lines with small squares).   Detections with a S/N of $<$\,5.0 are marked as upper limits.   The class II objects in the top and 
middle panels are among the 22 
distinguished using their W1-W2 (3.4\,$\mu$m-4.6\,$\mu$m) colour alone.   
The middle panels show the 2 objects that also have the large J-K excess, as well as having the brightest W3 signals, bright W4 signals 
and significant photometric variability (see Table 1 ).   
The class II objects in the bottom panels are 2 of the 5 with little 
or no W1-W2 excess that are included in the class II group on the basis of their bright W3 and/or W4 signals.}
\end{figure*}

\bsp

\label{lastpage}


\begin{thebibliography}{99}


\bibitem[\protect\citeauthoryear{Ardila et al.}{2000}]{ard00} Ardila D., Martin E., Basri G., 2000, AJ, 120, 479.







\bibitem[\protect\citeauthoryear{Carpenter et al.}{2006}]{car06} Carpenter J.M., Mamajek E.E., Hillenbrand L.A., Meyer M.R., 2006, ApJ, 651, L49.

\bibitem[\protect\citeauthoryear{Calvet et al.}{2002}]{cal02} Calvet N., D'Alessio P., Hartmann L., Wilner D., Walsh A., Sitko M., 2002, ApJ, 568, 1008C.


\bibitem[\protect\citeauthoryear{Chabrier et al.}{2000}]{cbah00} Chabrier G., Baraffe I., Allard F., Hauschildt P., 2000, ApJ, 542, 464.

\bibitem[\protect\citeauthoryear{Cutri et al.}{2011}]{cut12} Cutri R.M., Wright E.L., Conrow T., Bauer J., Benford D., Brandenburg H., Dailey J., Eisenhardt P.R.M., Evans T., Fajardo-Acosta S., Fowler J., Gelino C., Grillmair C., Harbut M., Hoffman D., Jarrett T., Kirkpatrick J.D., Leisawitz D., Liu W., Mainzer A., Marsh K., Masci F., McCallon H., Padgett D., Ressler M.E., Royer D., Shrutskie M.F., Stanford S.A., Wyatt P.L., Tholen D., Tsai C.W., Wachter S., Wheelock S.L., Yan L., Alles R., Beck R., Grav T., Masiero J., McCollum B., McGehee P., Papin M., Wittman M., 2012, Explanatory Supplement to the WISE All Sky Data Release.

\bibitem[\protect\citeauthoryear{Damjanov et al.}{2007}]{dam07} Damjanov I., Jayawardhana R., Scholz A., Ahmic M., Nguyen D.C., Brandeker A., van Kerkwijk M.H., 2007, ApJ, 670, 1337D.

\bibitem[\protect\citeauthoryear{Dawson, Scholz \& Ray}{2011}]{dsr11} Dawson P., Scholz A., Ray T.P., 2011, MNRAS, 418, 1231D.


\bibitem[\protect\citeauthoryear{de Bruijne et al.}{1997}]{deb97} de Bruijne J.H.J., Hoogerwerf R., Brown A.G.A., Aguilar L.A., de Zeeuw P.T., 1997, in Perryman M.A.C., Bernacca P.L., Battrick B., eds, in ESA SP-402: Hipparcos - Venice '97 Improved Methods for Identifying Moving Groups. p. 575.



\bibitem[\protect\citeauthoryear{Dullemond et al.}{2007}]{dul07} Dullemond C.P., Hollenbach D., Kamp. I., D'Alessio P., 2007, Protostars and Planets V. Univ. Arizona Press, Tucson, p. 555.

\bibitem[\protect\citeauthoryear{Dullemond \& Dominik}{2004}]{dul04} Dullemond C.P., Dominik C., 2004, A\&A, 421, 1075D.


\bibitem[\protect\citeauthoryear{Ercolano, Clarke \& Robitaille}{2009}]{ecr09} Ercolano B., Clarke C.J., Robitaille T.P., 2009, MNRAS, 394L, 141E.

\bibitem[\protect\citeauthoryear{Espaillat et al.}{2008}]{esp08} Espaillat C., Calvet N., Luhman K.L., Muzerolle J., D'Alessio P., 2008, ApJ, 682L, 125E.

\bibitem[\protect\citeauthoryear{Espaillat et al.}{2012}]{esp12} Espaillat C., Ingleby L., Hernandez J., Furlan E., D'Alessio P., Calvet N., Andrews S., Muzerolle J., Qi C., Wilner D., 2012, ApJ, 747, 103E.

\bibitem[\protect\citeauthoryear{Haisch et al.}{2001}]{hai01} Haisch K.E.jr., Lada E.A., Lada C.J., 2001, ApJ, 553, L153.



\bibitem[\protect\citeauthoryear{Hernandez et al.}{2007}]{her07} Hernandez J., Hartmann L., Megeath T., Gutermuth R., Muzerolle J., Calvet N., Vivas A.K., Brice{\~n}o C., Allen L., Stauffer J., 2007, ApJ, 662, 1067H.





\bibitem[\protect\citeauthoryear{Jayawardhana et al.}{2006}]{jay06} Jayawardhana R., Coffey J., Scholz A., Brandeker A., van Kerkwijk M.H., 2006, ApJ, 648, 1206J.



\bibitem[\protect\citeauthoryear{Lada et al.}{2006}]{lad06} Lada C.J., Muench A.A., Luhman K.L., Allen L., Hartmann L., Megeath T., Myers P., Fazio G., Wood K., Muzerolle J., Rieke G., Siegler N., Young E., 2006, AJ, 131, 1574.

\bibitem[\protect\citeauthoryear{Lawrence et al.}{2007}]{lawrence07} Lawrence A., Warren S.J., Almaini O., Edge A.C., Hambly N.C., Jameson R.F., Lucas P., Casali M., Adamson A., Dye S., Emerson J.P., Foucaud S., Hewett P., Hirst P., Hodgkin S.T., Irwin M.J., Lodieu N., McMahon R.G., Simpson C., Smail I., Mortlock D., Folger M., 2007, MNRAS, 379, 1599.


\bibitem[\protect\citeauthoryear{Lodieu et al.}{2006}]{lodieu06} Lodieu N., Hambly N.C., Jameson R.F., 2006, MNRAS, 373, 95.


\bibitem[\protect\citeauthoryear{Lodieu et al.}{2007}]{lodieu07} Lodieu N., Hambly N.C., Jameson R.F., Hodgkin S.T., Carraro G., Kendall T.R.,  2007, MNRAS, 374, 372.

\bibitem[\protect\citeauthoryear{Lodieu et al.}{2008}]{lodieu08} Lodieu N., Hambly N.C., Jameson R.F., Hodgkin S.T., 2008, MNRAS, 383, 1385.

\bibitem[\protect\citeauthoryear{Lodieu et al.}{2011}]{lodieu11} Lodieu N., Dobbie P.D., Hambly N.C., 2011, A\&A, 527A, 24L.


\bibitem[\protect\citeauthoryear{Luhman et al.}{2003}]{luh03} Luhman K.L., Stauffer J.R., Muench A.A., Rieke G.H., Lada E.A., Bouvier J., Lada C.J., 2003, ApJ, 593, 1093L.

\bibitem[\protect\citeauthoryear{Luhman}{2004}]{luh04a} Luhman K.L., 2004, ApJ, 602, 816L.

\bibitem[\protect\citeauthoryear{Luhman et al.}{2005}]{luh05} Luhman K.L., Lada C.J., Hartmann L., Muench A.A., Megeath S.T., Allen L.E., Myers P.C., Muzerolle J., Young E., Fazio G.G., 2005, ApJ, 631L, 69L.

\bibitem[\protect\citeauthoryear{Luhman}{2007}]{luh07a} Luhman K.L., 2007, ApJS, 173, 104L.

\bibitem[\protect\citeauthoryear{Luhman \& Mamajek}{2012}]{lam12} Luhman K.L., Mamajek E.E., 2012, ApJ, 758, 31L.




\bibitem[\protect\citeauthoryear{Martin et al.}{2004}]{mar04} Martin E.L., Delfosse X., Guieu S., 2004, AJ, 127, 449.

\bibitem[\protect\citeauthoryear{Mayne et al.}{2007}]{may07} Mayne N.J., Naylor T., Littlefair S.P., Saunders E.S., Jeffries R.D., 2007, MNRAS, 375, 1220M.

\bibitem[\protect\citeauthoryear{Merin et al.}{2010}]{mer10} Merin, B., Brown J.M., Oliveira I., Herczeg G.J., van Dishoeck E.W., Bottinelli S., Evans N.J.II., Cieza L., Spezzi L., Alcala J.N., Harvey P.M., Blake G.A., Bayo A., Geers V.G., Lahuis F., Prusti T., Augereau J-C., Olofsson J., Walter F.M., Kuenley C., 2010, ApJ, 718, 1200M.


\bibitem[\protect\citeauthoryear{Muzerolle et al.}{2006}]{muz06} Muzerolle J., Adame L., D'Alessio P., Calvet N., Luhman K.L., Muench A.A., Lada C.J., Rieke G.H., Siegler N., Trilling D.E., Young E.T., Allen L., Hartmann L., Megeath S.T., 2006, ApJ, 643, 1003M.

\bibitem[\protect\citeauthoryear{Muzerolle et al.}{2010}]{muz10} Muzerolle J., Allen L., Megeath S.T., Hernandez J., Gutermuth R.A., 2010, ApJ, 708, 1107M.

\bibitem[\protect\citeauthoryear{Natta \& Testi}{2001}]{nat01} Natta A., Testi L., 2001, A\&A, 376L, 22N.

\bibitem[\protect\citeauthoryear{Natta et al.}{2002}]{nat02} Natta A., Testi L., Comerón F., D'Antona F., Baffa C., Comoretto G., Gennari S., 2002, A\&A, 393, 597.




\bibitem[\protect\citeauthoryear{Oliveira et al.}{2002}]{olv02} Oliveira J.M., Jeffries R.D., Kenyon M.J., Thompson S.A., Naylor T.,  2002, A\&A, 382L, 22O.


\bibitem[\protect\citeauthoryear{Pecaut et al}{2012}]{pec12} Pecaut M.J., Mamajek E.E., Bubar E.J., 2012, ApJ, 746, 154P.

\bibitem[\protect\citeauthoryear{Preibisch \& Zinnecker}{1999}]{paz99} Preibisch T., Zinnecker H., 1999, AJ, 117, 2381P.

\bibitem[\protect\citeauthoryear{Preibisch et al.}{2002}]{pre02} Preibisch T., Brown A.G.A., Bridges T., Guenther E., Zinnecker H., 2002, AJ, 124, 404.



\bibitem[\protect\citeauthoryear{Riaz et al.}{2012}]{ria12} Riaz B., Lodieu N., Goodwin S., Stamatellos D., Thompson M., 2012, MNRAS, 420, 2497R.

\bibitem[\protect\citeauthoryear{Scholz et al.}{2007}]{sch07} Scholz A., Jayawardhana R., Wood K., Meeus G., Stelzer B., Walker C., O'Sullivan M., 2007, ApJ, 660, 1517.

\bibitem[\protect\citeauthoryear{Scholz et al.}{2009}]{sch09a} Scholz A., Xu X., Jayawardhana R., Wood K., Eisl\"offel J., Quinn C., 2009, MNRAS, 398, 873S.


\bibitem[\protect\citeauthoryear{Sherry et al.}{2004}]{she04} Sherry W.H., Walter F.M., Wolk S.J., 2004, AJ, 128, 2316S.



\bibitem[\protect\citeauthoryear{Skrutskie et al.}{2006}]{skrutskie06} Skrutskie M.F., Cutri R.M., Stiening R., Weinberg M.D., Schneider S., Carpenter J.M., Beichman C., Capps R., Chester T., Elias J., Huchra J., Liebert J., Lonsdale C., Monet D.G., Price S., Seitzer P., Jarrett T., Kirkpatrick J.D., Gizis J., Howard E., Evans T., Fowler J., Fullmer L., Hurt R., Light R., Kopan E.L., Marsh K.A., McCallon H.L., Tam R., Van Dyck S., Wheelock S., 2006, AJ, 131, 1163.

\bibitem[\protect\citeauthoryear{Slesnick et al.}{2006}]{sle06} Slesnick C.L., Carpenter J.M., Hillenbrand L.A., 2006, AJ, 131, 3016.

\bibitem[\protect\citeauthoryear{Slesnick et al.}{2008}]{sle08} Slesnick C.L., Hillenbrand L.A., Carpenter J.M., 2008, ApJ, 688, 377S.







\bibitem[\protect\citeauthoryear{Wright et al.}{2010}]{wri10} Wright E.L., Eisenhardt P.R.M., Mainzer A.K., Ressler M.E., Cutri R.M., Jarrett T., Kirkpatrick J.D., Padgett D., McMillan R.S., Skrutskie M., Stanford S.A., Cohen M., Walker R.G., Mather J.C., Leisawitz D., Gautier T.N.III., McLean I., Benford D., Lonsdale C.J., Blain A., Mendez B., Irace W.R., Duval V., Liu F., Royer D., Heinrichsen I., Howard J., Shannon M., Kendall M., Walsh A.L., Larsen M., Cardon J.G., Schick S., Schwalm M., Abid M., Fabinsky B., Naes L., Tsai C.-W., 2010, AJ, 140, 1868W.


\bibitem[\protect\citeauthoryear{Zaptero Osorio et al.}{2002}]{zap02} Zapatero Osorio M. R., Bejar V.J.S., Pavlenko Ya., Rebolo R., Allende Prieto C., Martín E.L., García López R.J., 2002, A\&A, 384, 937Z.

\end{thebibliography}
\end{document}